\documentclass[10pt]{asme2ej}
\usepackage{amsmath,amssymb,amsfonts}
\usepackage{graphicx} 
\usepackage{epsfig}
\usepackage{fancyhdr}
\usepackage{helvet}
\usepackage[hyphens]{url}
\usepackage{multicol}
\usepackage{newtxmath}
\usepackage{overpic}
\usepackage{setspace}
\usepackage{tikz}

\usepackage[switch,mathlines]{lineno}

\usepackage{hyperref}   
\hypersetup{
	colorlinks=true,
	linkcolor=blue,
	citecolor=blue,
	urlcolor=blue,
}
\usepackage[square,numbers]{natbib}

\newcommand\Rey{\mbox{\textit{Re}}}  
\newcommand\Mach{\mbox{\textit{Ma}}} 
\newcommand\Stokes{\mbox{\textit{St}}} 

\usetikzlibrary{calc}
\usetikzlibrary{arrows.meta}
\newcommand\Larrow[1]{\mathrel{\raisebox{0.4mm}{\!\!
			\begin{tikzpicture}[>=stealth]
				\node[inner sep=1ex] (a) {$\scriptstyle #1$};
				\draw[<-, line width = 0.8pt] ($(a.south west)+(0,0.05045)$) --($(a.south east)+(0,0.05045)$);
\end{tikzpicture}}}}

\title{Direct numerical simulation of particle-laden flow in a linear compressor cascade: Unsteady boundary-layer effects on particle-blade interactions}

\author{Taiyang Wang
	\affiliation{HEDPS, CAPT,\\
		School of Mechanics and Engineering Science,\\
		Peking University\\
		Beijing 100871, China\\
		Email: wangtaiyang@stu.pku.edu.cn
	}	
}

\author{Yaomin Zhao\thanks{Address all correspondence to this author.}
	\affiliation{HEDPS, CAPT,\\
		School of Mechanics and Engineering Science,\\
		Peking University\\
		Beijing 100871, China\\
		Email: yaomin.zhao@pku.edu.cn
	}	
}

\begin{document}

\maketitle    
\doublespacing
\begin{abstract}
{\it 
We perform point-particle direct numerical simulations (PP-DNS) of particle-laden flow through a linear compressor cascade subjected to synthetic free-stream turbulence.
Monodisperse particles are advanced in a one-way coupled Eulerian-Lagrangian framework with drag-only dynamics. 
We use PP-DNS collision statistics together with empirical deposition and erosion models to estimate particle deposition and blade erosion.
Collision hotspots and regions of predicted deposition are identified near the leading edge and over the pressure side. 
On the pressure side, for the intermediate Stokes number, the onset of frequent collisions and predicted deposition is spatially associated with elevated boundary-layer intermittency during bypass transition.
Nevertheless, for the largest particles, impacts occur farther upstream. 
On the suction side, sparse collisions and predicted deposition events appear only for the smallest particles and are phase-modulated by separation-induced vortex shedding. 
Joint distributions of impact velocity and angle show that leading-edge impacts are faster and span wider angles than pressure-side impacts.
For the two larger particle sizes, the higher normal impact velocities near the leading edge result in a greater likelihood of rebound than on the pressure side.
Therefore, appreciable erosion is predicted primarily near the leading edge under the present conditions.
The present results highlight the role of unsteady boundary-layer dynamics in affecting particle-blade interactions in compressor cascades.
}
\end{abstract}

\begin{nomenclature}
\subsection*{Symbols}
\entry{$\rho_{\mathrm{f}}$, $T_{\mathrm{f}}$}{fluid density and temperature}
\entry{$\mu_{\mathrm{f}}$, $\nu_{\mathrm{f}}$}{molecular and kinematic viscosity}
\entry{$U_{\mathrm{in}}$}{inlet velocity}
\entry{$\tau_{\mathrm{w}}$}{wall shear stress}
\entry{$\omega_z$}{spanwise vorticity}
\entry{$C_{\mathrm{ax}}$}{axial chord length}
\entry{$h$, $L_z$}{blade pitch and the spanwise domain size}
\entry{$\tau_\mathrm{f}$}{characteristic flow time}
\entry{$C_p$, $C_f$}{wall-pressure and wall-friction coefficient}
\entry{$\Rey_{\mathrm{f}}$, $\Mach_{\mathrm{f}}$}{Reynolds and Mach number}
\entry{$Tu$, $Ls$}{turbulence intensity and integral length scale}
\entry{$\gamma$}{intermittency}
\entry{$\varGamma$, $D$}{logical indicator and detector function}
\entry{$\varPhi_\mathrm{p}$}{volume fraction of the solid phase}
\entry{$\rho_{\mathrm{p}}$, $d_{\mathrm{p}}$, $V_{\mathrm{p}}$, $m_{\mathrm{p}}$}{particle density, size, volume and mass}
\entry{$\boldsymbol{r}_{\mathrm{p}}$, $\boldsymbol{u}_{\mathrm{p}}$}{particle position and velocity vectors}
\entry{$\tau_\mathrm{p}$}{particle relaxation time}
\entry{$C_{\mathrm{d}}$}{drag coefficient}
\entry{$\Rey_{\mathrm{p}}$}{particle Reynolds number}
\entry{$\Stokes$}{Stokes number}
\entry{$s, n$}{streamwise and wall-normal coordinates}
\entry{$\mathrm{CoR}_n$}{normal coefficient of restitution}
\entry{$W_{\mathrm{cont}}$}{the work required to overcome adhesion forces}
\entry{$E_{\mathrm{crit}}$}{maximum stored elastic energy}
\entry{$w_{\mathrm{crit}}$, $w_{\mathrm{max}}$}{critical deformation and maximum deformation}
\entry{$\gamma_s$}{surface free energy}
\entry{$A_{\mathrm{cont}}$, $A_{\mathrm{crit}}$}{contact surface area and surface area at maximum elastic deformation}
\entry{$E_\mathrm{p}$, $E_\mathrm{b}$}{Young's moduli of the particle and the blade}
\entry{$\nu_\mathrm{p}$, $\nu_\mathrm{b}$}{Poisson's ratios of the particle and the blade}
\entry{$E_c$}{composite modulus}
\entry{$\sigma_y$}{yield stress}
\entry{$u_{\mathrm{p}1}$, $\beta_1$}{impact velocity and angle}
\entry{$\varepsilon$, $E$}{erosion rate and erosion rate density}
\end{nomenclature}

\section{Introduction}\label{introduction}

Solid particles pose a significant threat to gas turbine engines, primarily
through deposition and erosion of engine components \cite{Grant1975Erosion,Rozati2010Effects}.
Particle deposition can lead to fouling, increasing blade surface roughness
and potentially altering blade geometry, whereas erosion primarily results in
material removal and degradation of structural integrity.
These effects can reduce the lifetime and operational efficiency of engine components, potentially leading to catastrophic failure \cite{Dunn2012Operation}. 
Therefore, understanding the effects of particle deposition and erosion represents a critical challenge in turbomachinery design.

Early studies on particle erosion in turbomachinery flows were mostly based on experiments.
Grant and Tabakoff \cite{Grant1975Erosion} studied the erosion profiles in a one-and-a-half-stage compressor, finding that the leading edge of the rotor undergoes severe erosion.
Nevertheless, the photographic methods used in that study limited observations to particles larger than $30~\mu\mathrm{m}$ \cite{Hamed2006Erosion}.
Using a laser Doppler velocimeter system, Tabakoff \textit{et al.} \cite{Tabakoff1987Laser} investigated the erosive impact of smaller particles, observing that the blade erosion depends significantly on the impact angle.
Later, Tabakoff \cite{Tabakoff1987Study} experimentally examined the performance deterioration of a single-stage axial compressor resulting from erosion.
It was concluded that the efficiency loss was due to the deterioration of the leading edges as well as changes in the airfoil shape.
This conclusion was also confirmed in the subsequent study \cite{Ghenaiet2004Experimental}.
Further evidence of performance deterioration was provided through a series of full-scale engine tests \cite{Dunn1987Performance,Dunn2012Operation}.

The aforementioned erosion studies primarily focused on particles with diameters larger than $10~\mu\mathrm{m}$.
Previous studies have shown that smaller particles tend to follow the flow streamlines closely, resulting in relatively low impact efficiencies and generally weaker erosive effects.
Nevertheless, once these particles impact a surface, they are more likely to adhere
than to rebound, since adhesive forces become increasingly dominant as particle size decreases \cite{Dahneke1975Further,Wall1990Measurements}.
Focusing on turbomachinery flows, numerous experiments have shown that particles deposit predominantly near the leading edge and on the pressure side \cite{Tabakoff1987Laser,Dunn2012Operation,Webb2012Coal}, whereas the specific contribution of smaller particles remains difficult to isolate.
Moreover, these observations mainly characterize the final deposition pattern, whereas the underlying particle response to the local flow field remains insufficiently resolved.
In fact, limited by current experimental methods, it remains challenging to obtain high-quality simultaneous measurements of the fluid and particle phases \cite{Ghenaiet2024Modeling}.

Numerical simulations have become an increasingly important tool for particle-laden flow research due to rapid advancements in computational methods and hardware.
For simulations of particle-laden flows, two frameworks are typically employed: one is the Eulerian-Eulerian framework, and the other is the Eulerian-Lagrangian framework \cite{Balachandar2010Turbulent}.
The Eulerian-Eulerian framework, in which both gas and solid phases are treated as continua and the two-fluid model is employed \cite{Baer1986two}, is often criticized for its lack of accuracy, primarily due to its oversimplification of the solid phase and strong reliance on constitutive closures \cite{Slater2001calculation,Slater2003Particle,Sundaresan2018Toward}.
In the Eulerian-Lagrangian approach, however, the fluid phase is solved within an Eulerian framework, while solid particles are tracked in a Lagrangian manner \cite{Brandt2022Particle}. 
Specifically, the forces acting on the particles are determined through fluid-particle interactions, and particle dynamics are governed by Newton's equations of motion \cite{Guha2008Transport}.
The Eulerian-Lagrangian framework has become the predominant method for investigating gas-particle flows, and numerous numerical simulations have been performed for cases including homogeneous isotropic turbulence \cite{Yeung1989Lagrangian}, planar channel flow \cite{Jie2022On}, and flat-plate boundary-layer flow \cite{Xiao2020Eulerian,Chen2022Two,Yu2024Transport}.

Under the Eulerian-Lagrangian framework, different strategies can be used for both the gas phase and the particle phase.
Regarding the particle phase, most existing studies utilize the point-particle approximation, in which particles are typically treated as constant-density spheres and are assumed to be smaller than the Kolmogorov length scale \cite{Balachandar2010Turbulent, Brandt2022Particle}.
Compared to particle-resolved simulations in which the flow fields around particles need to be fully resolved, simulations with point particles require far fewer computational resources. 
For the gas phase, the flow field can be calculated using Reynolds-averaged Navier--Stokes (RANS), large-eddy simulation (LES) or direct numerical simulation (DNS).
However, the turbulence models introduced in RANS and LES may cause unphysical effects \cite{Crowe2011Multiphase}, casting doubt on their applicability when simulating complex phenomena such as flow separation and transition.
In comparison, point-particle direct numerical simulation (PP-DNS), in which the turbulent flow is fully resolved rather than modeled, has proved valuable for investigating the statistical behavior of particles \cite{Guha2008Transport, Brandt2022Particle}. 
Nevertheless, existing PP-DNS studies mainly focus on canonical cases with simple geometries, and applications to turbomachinery geometries remain scarce.

Focusing on turbomachinery erosion, Tabakoff and coworkers conducted a number of pioneering simulations \cite{Hussein1973Dynamic, Elfeki1987Erosion}.
Because the volume fraction of the solid phase $\varPhi_\mathrm{p}$ in engine-relevant gas-solid mixtures is typically below $10^{-6}$ \cite{Elghobashi1994On}, their studies adopted the one-way coupling strategy, which neglects the feedback of the solid phase on the fluid phase \cite{Brandt2022Particle}. 
Nevertheless, these early studies were restricted to two-dimensional scenarios and treated the flow as inviscid.
Subsequently, most studies used three-dimensional RANS \cite{Ghenaiet2012Study, Vogel2018Simulation, Ghenaiet2024Modeling}, and employed a partitioned approach: a time-averaged flow field was first computed, after which particles were propagated through it.
In addition to erosion studies, this method has also been applied to particle-deposition modeling \cite{Barker2012Coal,Prenter2016Deposition}.
Although this strategy significantly reduces computational resource requirements, the steady assumption of turbomachinery flows is often unrealistic \cite{Sandberg2022Fluid}.
This shortcoming was addressed by Beck \textit{et al.} \cite{Beck2019Towards}, who performed the first LES of particle-laden flow in a turbomachinery configuration.
However, LES employs turbulence models for subgrid-scale fluctuations, and thus particle responses to unresolved turbulence cannot be captured directly \cite{Armenio1999Effect, Kuerten2004Can}.
Therefore, to enable a deeper investigation of particle behavior in practical turbomachinery applications, simulations that resolve all turbulent scales are necessary.
However, to the best of the authors' knowledge, PP-DNS databases for particle-laden turbomachinery flows remain scarce.

In the present study, we perform PP-DNS of particle-laden flow through a linear compressor cascade subjected to controlled free-stream turbulence. 
The fully resolved blade boundary layers enable a mechanistic assessment of how unsteady phenomena, \emph{i.e.} pressure-side bypass transition and suction-side separation-induced shedding, affect particle-blade interactions. 
We consider monodisperse particles at three Stokes numbers within a nominal point-particle framework and quantify (i) collision locations and frequencies, (ii) their relationship with boundary-layer intermittency and shedding dynamics, (iii) deposition patterns predicted using the Bons model \cite{Bons2017Simple}, and (iv) leading-edge erosion predicted using the Grant--Tabakoff erosion model \cite{Grant1975Erosion}. 
This study provides a systematic PP-DNS comparison of particle-blade interactions across multiple Stokes numbers in a compressor cascade.

\section{Numerical Simulation}\label{methodology}

\begin{figure*}[!t]
	\centering
	\begin{overpic}[width=0.95\textwidth]{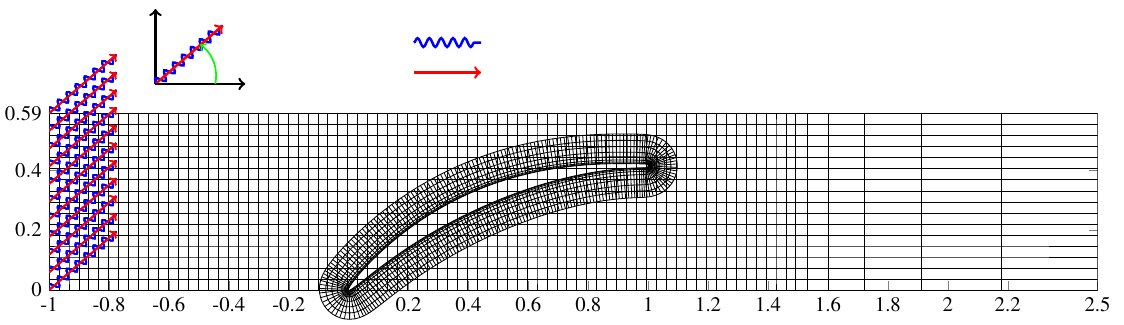}
		{
			\put(0,11){$y$}
			\put(50,-1){$x$}
			
			\put(20,23){$\alpha=41^\circ$}
			\put(15,27){$y$}
			\put(23,21){$x$}
			
			\put(44,21.75){inlet velocity $U_\mathrm{in}$}
			\put(44,24.5){synthetic turbulence $(u^\prime,v^\prime,w^\prime)$}
		}
	\end{overpic}
	\caption{Schematic of the axial and pitchwise $(x-y)$ plane of the particle-laden flow in a linear compressor cascade. The computational grid consists of a background H-type grid and an O-type grid, with approximately every sixteenth grid line shown in each coordinate direction.}
	\label{setup}
\end{figure*}
\begin{table}[!b]
	\caption{Grid parameters.}
	\begin{center}
		\label{grid_parameter}
		\begin{tabular}{c c c c c c}
			\hline\hline
			Grid & $N_x$ & $N_y$ & $L_z$ & $N_z$ & Points \\
			\hline
			H-type grid & $1540$ & $337$ & $0.24$ & $160$ & $83,036,800$\\
			O-type grid & $2152$ & $137$ & $0.24$ & $160$ & $47,171,840$\\
			\hline\hline
		\end{tabular}
	\end{center}
\end{table}

The primary objective of this study is to use PP-DNS collision statistics, together with empirical deposition and erosion models, to assess the effects of sand particles of different sizes on a compressor-blade surface.
For simplicity, the volume fraction of sand particles is assumed to be sufficiently low to adopt the one-way coupling strategy, and thus the solid phase is driven by the gas phase without feedback \cite{Brandt2022Particle}.
The numerical treatment and setup for both phases are described in detail below.

\subsection{Gas Phase}

For the flow simulation, a linear compressor cascade with the NACA 65 blade profile is assumed to operate at room temperature.
Based on the inlet velocity, the operating axial Reynolds number is $\Rey_{\mathrm{f}}=138,500$ and the Mach number is $\Mach_{\mathrm{f}}=0.15$, corresponding to the previous work of Zaki \textit{et al.} \cite{Zaki2010Direct}.
The resulting unsteady flow field is obtained by solving the three-dimensional compressible Navier--Stokes equations using the solver HiPSTAR \cite{Sandberg2015Compressible}.

Figure~\ref{setup} shows a schematic of the numerical setup.
The use of the overset grid method \cite{Deuse2020Implementation} requires a hybrid H-O-type grid, whose detailed parameters are provided in Table~\ref{grid_parameter}.
The present DNS is conducted in a Cartesian coordinate system $(x,y,z)$, where the coordinates $x$, $y$ and $z$ correspond to the axial, pitchwise, and spanwise directions, respectively.
The background H-type grid governs the computational domain, which spans one blade pitch $h=0.59C_\mathrm{ax}$ in the pitchwise direction and has a spanwise extent of $L_z=0.24C_\mathrm{ax}$.
Here, $C_\mathrm{ax}=1$ is the non-dimensional axial chord length. 
Periodic boundary conditions are applied in both the pitchwise and spanwise directions.
The O-type grid ensures accurate resolution of the blade boundary layer, with an adiabatic, no-slip blade surface.
At the inlet, a uniform velocity condition $(U_\mathrm{in}\cos(\alpha),U_\mathrm{in}\sin(\alpha),0)$ is imposed, as indicated by the red arrow in Fig.~\ref{setup}.
Here, $U_\mathrm{in}=1$ is the non-dimensional inlet velocity and $\alpha=41^\circ$ is the inlet flow angle.
To replicate realistic incoming turbulence, synthetic free-stream turbulence (FST) symbolized by the blue wavy line is generated at the inlet using a digital filter approach \cite{Klein2003filter, Touber2009large}.
The turbulence intensity is $Tu=3.25\%U_\mathrm{in}$ and the integral length scale is $Ls=0.06C_\mathrm{ax}$, the same as those used in the previous study \cite{Zaki2010Direct}.
In addition, a characteristic boundary condition combined with a sponge layer is applied at the outlet to prevent spurious reflections \cite{Sandberg2006Nonreflecting}.

In the present cases, the grid spacings around the blade surface, normalized by the viscous scale, are below $(9, 0.8, 10)$ in the streamwise, wall-normal and spanwise directions, respectively.
The resolution is similar to that used in previous DNS studies \cite{Zaki2010Direct} and is considered sufficient to resolve the blade boundary layer.
Thus, the present DNS flow fields can serve as the basis for subsequent Lagrangian particle tracking.

\subsection{Solid Phase}
For the solid phase, the physical models, numerical strategies and parameter setup are discussed below.
The sand particles are modeled as point particles, idealized as rigid spheres with a uniform diameter $d_{\mathrm{p}}$ and density $\rho_{\mathrm{p}}$.
The non-dimensional equations of motion for these Lagrangian particles are given by \cite{Chen2022Two}
\begin{equation}
	\begin{aligned}
		\frac{\mathrm{d}\boldsymbol{r}_{\mathrm{p}}}{\mathrm{d}t}&=\boldsymbol{u}_{\mathrm{p}},\\
		m_{\mathrm{p}}\frac{\mathrm{d}\boldsymbol{u}_{\mathrm{p}}}{\mathrm{d}t}&=\boldsymbol{f}_{\mathrm{drag}},
	\end{aligned}
	\label{particle_equations}
\end{equation}
where $\boldsymbol{r}_{\mathrm{p}}$ and $\boldsymbol{u}_{\mathrm{p}}$ represent the position and velocity vectors of each particle, respectively.
$m_{\mathrm{p}}=\rho_{\mathrm{p}}V_{\mathrm{p}}$ is the particle mass, where $V_{\mathrm{p}}=\pi d_{\mathrm{p}}^3/6$ denotes the particle volume.
Owing to the significantly higher density of the solid phase compared to the gas phase, only the drag force $\boldsymbol{f}_{\mathrm{drag}}$ \cite{Elghobashi1992} is considered, given by
\begin{equation}
	\boldsymbol{f}_{\mathrm{drag}}=3\pi\mu_{\mathrm{f}}d_{\mathrm{p}}\frac{C_{\mathrm{d}}}{24}\frac{\Rey_{\mathrm{p}}}{\Rey_{\mathrm{f}}}(\boldsymbol{u}_{\mathrm{f}}-\boldsymbol{u}_{\mathrm{p}}),
\end{equation}
where $\boldsymbol{u}_{\mathrm{f}}$ represents the fluid velocity at the particle's center of mass.
Here, $\mu_{\mathrm{f}}$ is the fluid dynamic viscosity, which depends on the local temperature $T_\mathrm{f}$ via Sutherland's law \cite{White1991viscous}
\begin{equation}
	\mu_\mathrm{f}=T^{3/2}_\mathrm{f}\frac{1+C_\mathrm{Suth}}{T_\mathrm{f}+C_\mathrm{Suth}}.
\end{equation}
Here, $C_\mathrm{Suth}=0.3766$ is Sutherland's constant corresponding to the room temperature of $293.15~\mathrm{K}$.
Moreover, $\Rey_\mathrm{p}=\Rey_\mathrm{f}|\boldsymbol{u}_\mathrm{f}-\boldsymbol{u}_\mathrm{p}|d_\mathrm{p}/\nu_\mathrm{f}$ is the particle Reynolds number, where $\nu_\mathrm{f}$ is the fluid kinematic viscosity.
In addition, $C_{\mathrm{d}}$ is the drag coefficient \cite{Schiller1935drag}, which is calculated as
\begin{equation}
	C_{\mathrm{d}}=\left\{
	\begin{aligned}
		&\dfrac{24}{\Rey_{\mathrm{p}}}\left(1+0.15\Rey_{\mathrm{p}}^{0.687}\right)   &\Rey_{\mathrm{p}}\le1000,\\
		&0.44                                                                        &\Rey_{\mathrm{p}}>1000.
	\end{aligned}\right.
\end{equation}
Note that particles may also experience the Saffman lift force $\boldsymbol{f}_{\mathrm{saff}}$ arising from strong local shear \cite{Saffman1965lift}.
An a posteriori evaluation over 120 instantaneous snapshots yielded ensemble-mean ratios $\langle|\boldsymbol{f}_{\mathrm{saff}}|\rangle/\langle|\boldsymbol{f}_{\mathrm{drag}}|\rangle$ of $1.4\times10^{-3}$, $9.2567\times10^{-4}$, and $3.4274\times10^{-4}$ for cases d1, d2, and d4, respectively.
This diagnostic is based on drag-only trajectories and does not exclude larger local ratios for a small subset of near-wall particles. 
Nevertheless, it indicates that Saffman lift is secondary on an ensemble-average basis under the present conditions and is therefore neglected.

The particle equations of motion (\ref{particle_equations}) are integrated using the velocity Verlet algorithm
\cite{Tartakovsky2005Modeling, Tartakovsky2016Pairwise}:
\begin{equation}
	\begin{aligned}
		\boldsymbol{r}_\mathrm{p}(t+\Delta t)&=\boldsymbol{r}_\mathrm{p}(t)+\Delta t\boldsymbol{u}_\mathrm{p}+\frac{\Delta t^2}{2m_\mathrm{p}}\boldsymbol{f}_{\mathrm{drag}}(t),\\
		\boldsymbol{u}_\mathrm{p}(t+\Delta t)&=\boldsymbol{u}_\mathrm{p}(t) +\frac{\Delta t}{2m_\mathrm{p}}\left[\boldsymbol{f}_{\mathrm{drag}}(t)+\boldsymbol{f}_{\mathrm{drag}}(t+\Delta t) \right],\\
	\end{aligned}
\end{equation}
where $\Delta t$ is the time step.
Here, $\boldsymbol{f}_{\mathrm{drag}}(t+\Delta t)$ represents the drag force computed based on the particle position at time $(t+\Delta t)$, and the required flow variables at the particle location are obtained through trilinear interpolation \cite{Xiao2020Eulerian, Yu2024Transport}.
A particle-blade collision is detected when the particle center enters the blade geometry \cite{Norouzi2016Coupled,Beck2019Towards}.
During trajectory integration, particle rebound is represented by an ideal specular-reflection condition relative to the local tangent plane, with the finite particle size neglected.
Note that this condition conserves impact kinetic energy and does not represent physical adhesion or dissipative rebound.
The Bons deposition criterion \cite{Bons2017Simple} described in \S~\ref{Particle_Deposition} is therefore applied in post-processing to first-collision events only, and no physical interpretation is assigned to post-impact trajectories or repeated impacts.

Unlike most particle-laden simulations conducted on a single rectangular grid, the present simulations use a hybrid overlapping H-O-type grid.
This topology and geometry pose a significant challenge for efficient trajectory tracking of numerous particles, particularly within the curved O-type grid.
To address this, we employ the framework based on overlapping grids \cite{wang2025high}, thereby enabling large-scale direct numerical simulations.
Because this particle-tracking framework was developed within HiPSTAR, the same compressible solver is used in the present study despite the low inlet Mach number.
The accuracy and computational efficiency of the framework have been validated using a wide range of test cases, and further implementation details are provided in \cite{wang2025high}.

\begin{table}[!b]
	\caption{Parameters for particle-laden compressor flow cases.}
	\begin{center}
		\label{particle_parameter}
		\begin{tabular}{p{50pt}p{40pt}p{40pt}p{40pt}p{40pt}}
			\hline\hline
			\centering Case & 
			\centering $\rho_\mathrm{p}^*~(\mathrm{kg/m^3})$ & 
			\centering $d_\mathrm{p}^*~(\mu\mathrm{m})$ & 
			\centering $d^*_\mathrm{p}/\eta^*_\mathrm{f}$ & 
			\multicolumn{1}{c}{$\Stokes$} \\
			\hline
			\centering case d1 & 
			\centering $2690$ & 
			\centering $1$ &  
			\centering $0.2267$ & 
			\multicolumn{1}{c}{$0.0101$} \\
			\centering case d2 & 
			\centering $2690$ & 
			\centering $2$ &  
			\centering $0.4534$ & 
			\multicolumn{1}{c}{$0.0403$} \\
			\centering case d4 & 
			\centering $2690$ & 
			\centering $4$ &  
			\centering $0.9067$ &
			\multicolumn{1}{c}{$0.1612$} \\
			\hline\hline
		\end{tabular}
	\end{center}
\end{table}

The physical parameters of the solid phase are specified as follows.
The ingested sand particles are represented by calcia-magnesia-alumina-silicate (CMAS) particles \cite{Bansal2015Properties}, with a density of $\rho_{\mathrm{p}}^*=2690~\mathrm{kg/m^3}$.
To meet the commonly adopted criterion for the point-particle approximation \cite{Balachandar2010Turbulent}, the particle diameters $d_{\mathrm{p}}^*$ are chosen to be smaller than the Kolmogorov scale $\eta^*_\mathrm{f}\approx4.5~\mathrm{\mu m}$.
This threshold corresponds to particle sizes that can evade upstream particle separators \cite{Tabakoff1987Laser,Dunn2012Operation,Webb2012Coal} and are more challenging to characterize experimentally than their larger counterparts.
Accordingly, three particle-size cases are considered, and detailed parameters are summarized in Table~\ref{particle_parameter}.
Note that for case d4, the ratio $d^*_{\mathrm p}/\eta^*_{\mathrm f}=0.9067$ lies close to the upper limit of the point-particle regime, and finite-size and near-wall effects may be more relevant than in cases d1 and d2.
Moreover, the Stokes number $\Stokes=\tau_\mathrm{p}/\tau_\mathrm{f}$, defined as the ratio of the particle relaxation time $\tau_\mathrm{p}$ to a characteristic flow time $\tau_\mathrm{f}$, is introduced to characterize particle behavior.
Here, $\tau_\mathrm{p}$ is calculated as $\tau_\mathrm{p}=\Rey_{\mathrm{f}}\rho_\mathrm{p}d_\mathrm{p}^2/(18\mu_\mathrm{f})$, and $\tau_\mathrm{f}=C_\mathrm{ax}/U_\mathrm{in}$ is defined as the ratio of the axial chord length to the inlet velocity.

Furthermore, to simulate the continuous ingestion of sand, particles are injected into the computational domain.
To facilitate comparison of the particle-blade interactions, the same particle-number injection rate is imposed in all three cases.
Numerically, at every non-dimensional time interval $\Delta t_\mathrm{add}=0.012$, a total of $N_\mathrm{add}=1755$ particles are emitted from fixed positions, which are uniformly distributed at the inlet.
The initial particle velocity is set to match the local fluid velocity, obtained via interpolation from the surrounding flow field.
To reduce computational cost, particles that travel beyond $x>1.65$ are removed.
Under this setup, approximately $6\times10^5$ particles are present in the domain in the statistically stationary state, resulting in a solid-phase volume fraction of $\varPhi_\mathrm{p}<10^{-6}$, which satisfies the criterion for the one-way coupling strategy \cite{Elghobashi1994On}.
The entire sand-ingestion process lasts approximately $14C_\mathrm{ax}/U_\mathrm{in}$, and the total particle number reaches an approximately constant value at $t=4.8C_\mathrm{ax}/U_\mathrm{in}$.
The particle-blade collision events recorded over the remaining $9.2C_\mathrm{ax}/U_\mathrm{in}$ provide a sufficiently large sample for the subsequent deposition and erosion analyses.
Note that since the suction-side events in case d1 are rare, their detailed local distribution retains appreciable statistical uncertainty.

\section{Results}
\subsection{Overview of the Particle-Laden Flow Field}

\begin{figure}[!t]
	\centering\small
	\begin{overpic}[width=0.49\textwidth]{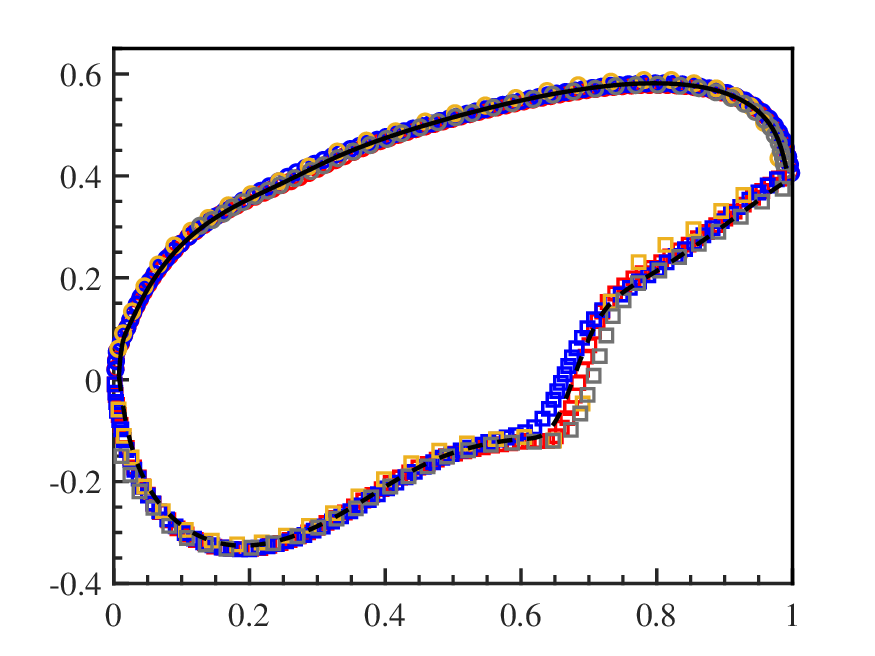}
		{
			\put(1,68){$(a)$}
			\put(3,38){$C_p$}
			\put(50.5,0){$x$}
		}
	\end{overpic}
	\begin{overpic}[width=0.49\textwidth]{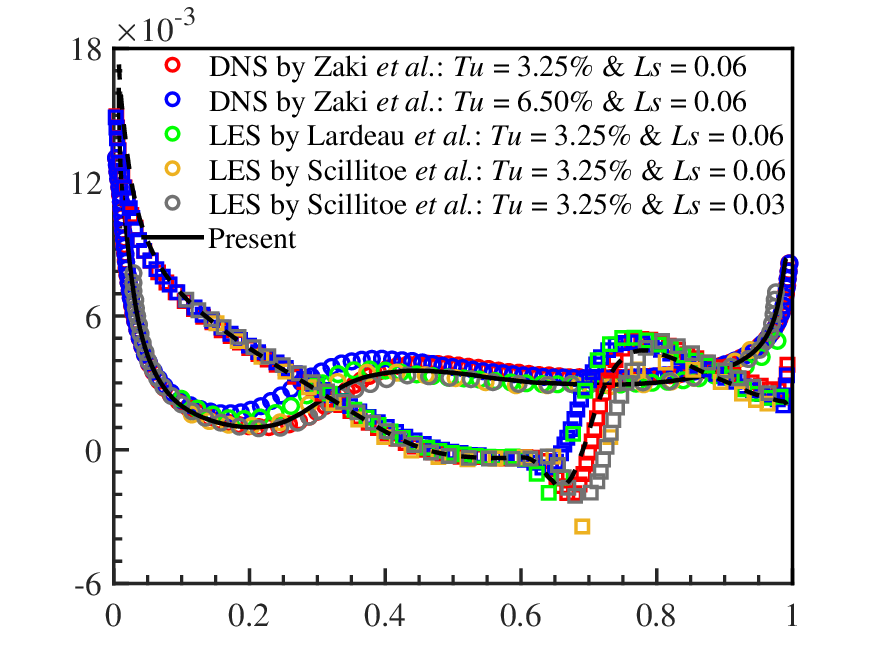}
		{
			\put(1,68){$(b)$}
			\put(3,38){$C_f$}
			\put(50.5,0){$x$}
		}
	\end{overpic}
	\caption{Statistics of the boundary layer around the blade surface: 
		$(a)$ wall-pressure coefficient $C_p$; $(b)$ wall-friction coefficient $C_f$. 
		Here, the results on the pressure side are denoted by circles and solid lines, while the results on the suction side are denoted by squares and dashed lines.}
	\label{validation} 
\end{figure}

We first present the mean statistics of the blade boundary layer in Fig.~\ref{validation}.
The pressure coefficient is defined as $C_p=(p_\mathrm{w}-p_0)/(0.5\rho_\mathrm{in}U_\mathrm{in}^2)$, where $p_\mathrm{w}$ denotes the wall pressure and $p_0$ is a reference pressure.
The wall-friction coefficient $C_f$ is defined as $C_f=\tau_{\mathrm{w}}/(0.5\rho_\mathrm{in}U_\mathrm{in}^2)$, where $\tau_{\mathrm{w}}$ is the wall shear stress.
The distributions of $C_p$ and $C_f$ around the blade are shown in Fig.~\ref{validation}$(a)$ and Fig.~\ref{validation}$(b)$, respectively.
On the pressure side, represented by the solid lines, the blade boundary layer experiences an adverse pressure gradient (APG) starting from the leading edge ($0<x\lesssim0.8$), followed by a favorable pressure gradient (FPG) at $0.8\lesssim x$.
In the APG region, $C_f$ increases from $x\approx0.2$ to $x\approx0.4$, suggesting a transition of the boundary layer from the laminar to the turbulent state.
Moreover, the pressure-side boundary-layer flow accelerates in the FPG region, and $C_f$ increases further accordingly at $0.8\lesssim x$.
On the suction side, the blade boundary layer first undergoes an FPG, which is then followed by a severe APG ($0.2\lesssim x$).
The strong APG causes the boundary layer to separate, forming a large separation region at $0.46\lesssim x\lesssim0.69$ as evidenced by the negative wall-friction coefficient $C_f<0$ in Fig.~\ref{validation}$(b)$.
Downstream of the separation, the flow reattaches following transition to turbulence, as suggested by the sudden increase of $C_f$ at $x\approx 0.7$.
Importantly, both the wall-pressure and wall-friction coefficients from the present simulations show good agreement with previous results \cite{Zaki2010Direct,Lardeau2012Large,Scillitoe2016Numerical,Scillitoe2019Large}, as indicated by the comparison between the lines and symbols.
The slight discrepancies among these cases, however, presumably result from differences among the methods used to generate the incoming FST.

\begin{figure}[!t]
	\centering\small
	\begin{overpic}[width=0.6\textwidth]{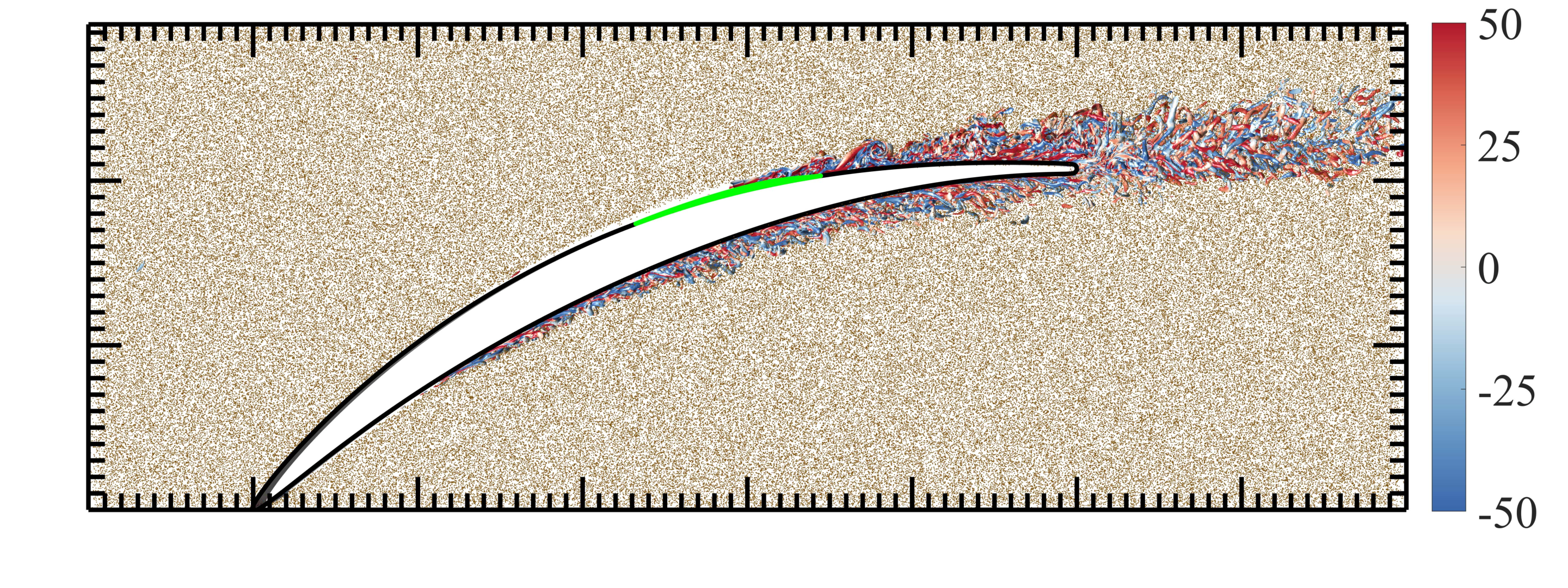}
		{
			\put(-1,34.5){$(a)$}
			\put(1,19.5){$y$}
			\put(2.25,4){0}
			\put(-0.5,14.5){0.2}
			\put(-0.5,25){0.4}
			\put(51.5,0.5){$x$}
			\put(2,1.5){-0.2}
			\put(15,1.5){0}
			\put(24,1.5){0.2}
			\put(34.5,1.5){0.4}
			\put(45,1.5){0.6}
			\put(55.5,1.5){1.0}
			\put(66,1.5){1.2}
			\put(76.5,1.5){1.4}
			\put(87,1.5){1.6}
			\put(96,19.5){$\omega_z$}
			
			\put(7,27){\rotatebox{-138.5}{$\Larrow{\rule{2cm}{0pt}}$}}
			\put(7,16){\rotatebox{41}{inflow direction}}
		}
	\end{overpic}
	\begin{overpic}[width=0.6\textwidth]{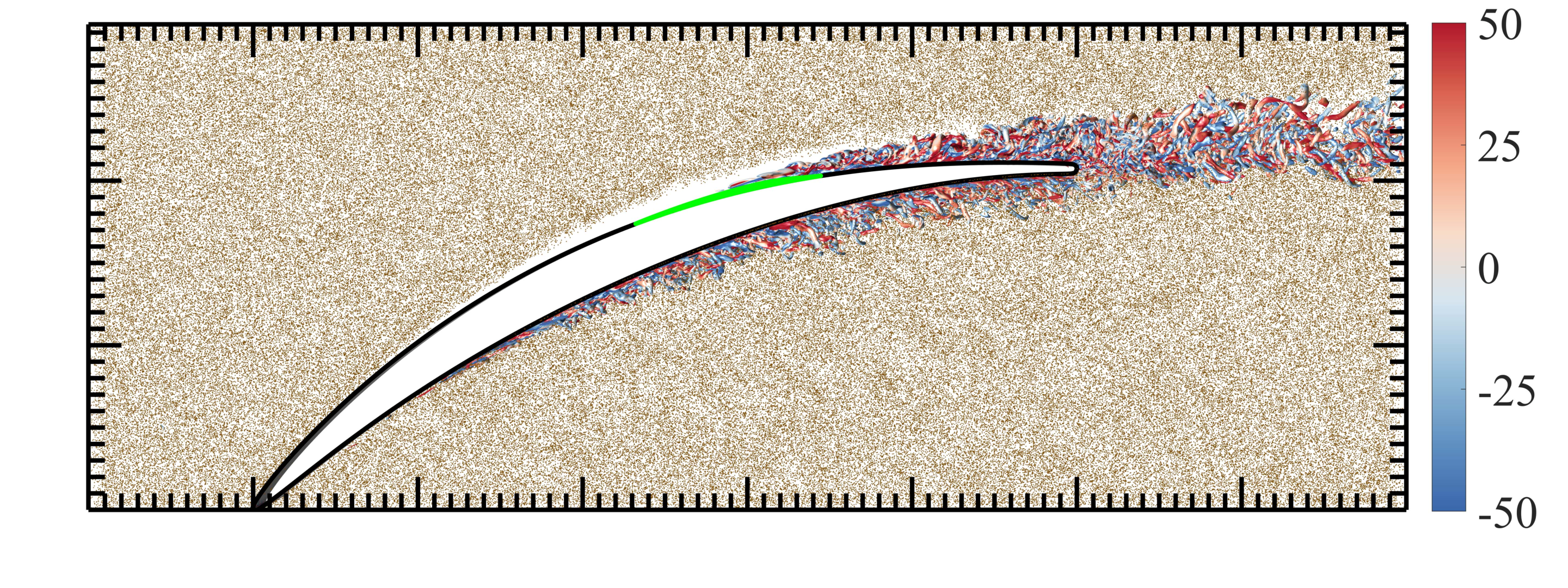}
		{
			\put(-1,34.5){$(b)$}
			\put(1,19.5){$y$}
			\put(2.25,4){0}
			\put(-0.5,14.5){0.2}
			\put(-0.5,25){0.4}
			\put(51.5,0.5){$x$}
			\put(2,1.5){-0.2}
			\put(15,1.5){0}
			\put(24,1.5){0.2}
			\put(34.5,1.5){0.4}
			\put(45,1.5){0.6}
			\put(55.5,1.5){1.0}
			\put(66,1.5){1.2}
			\put(76.5,1.5){1.4}
			\put(87,1.5){1.6}
			\put(96,19.5){$\omega_z$}
			
			\put(7,27){\rotatebox{-138.5}{$\Larrow{\rule{2cm}{0pt}}$}}
			\put(7,16){\rotatebox{41}{inflow direction}}
		}
	\end{overpic}
	\begin{overpic}[width=0.6\textwidth]{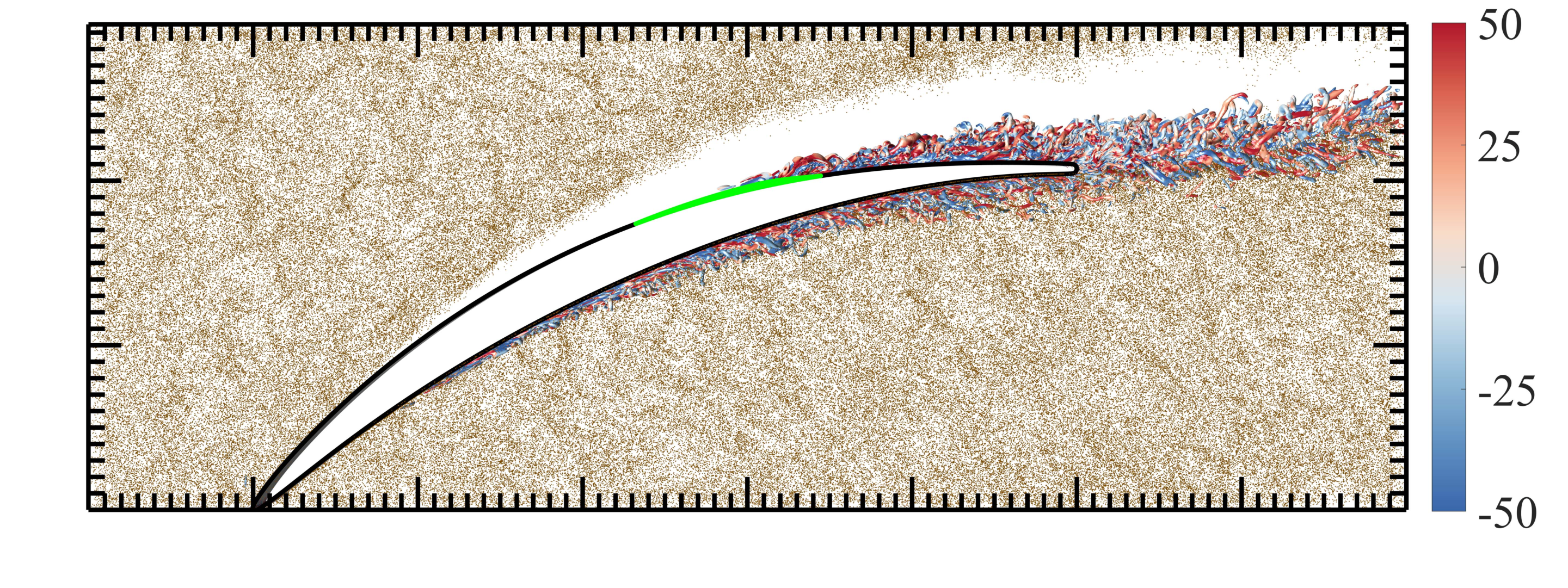}
		{
			\put(-1,34.5){$(c)$}
			\put(1,19.5){$y$}
			\put(2.25,4){0}
			\put(-0.5,14.5){0.2}
			\put(-0.5,25){0.4}
			\put(51.5,0.5){$x$}
			\put(2,1.5){-0.2}
			\put(15,1.5){0}
			\put(24,1.5){0.2}
			\put(34.5,1.5){0.4}
			\put(45,1.5){0.6}
			\put(55.5,1.5){1.0}
			\put(66,1.5){1.2}
			\put(76.5,1.5){1.4}
			\put(87,1.5){1.6}
			\put(96,19.5){$\omega_z$}
			
			\put(7,27){\rotatebox{-138.5}{$\Larrow{\rule{2cm}{0pt}}$}}
			\put(7,16){\rotatebox{41}{inflow direction}}
		}
	\end{overpic}
	\caption{An axial and pitchwise $(x-y)$ overview of $(a)$: case d1; $(b)$: case d2; $(c)$: case d4.
		Here, sand particles are represented by brown spheres, and vortical structures identified using the $Q$ criterion are colored according to the spanwise vorticity $\omega_z$.
		Moreover, the inflow direction is marked by the black arrow, and the mean separation region is denoted by the green line.}
	\label{xy-overview} 
\end{figure}

We further visualize the particle-laden flow fields.
Figure~\ref{xy-overview} provides an overview of the flow structures and particle distribution through an instantaneous snapshot in the $(x-y)$ plane for each case.
For the gas phase, the boundary-layer transition is clearly shown by the emergence of abundant vortical structures identified by the $Q$ criterion \cite{Hunt1988Eddies}, while the flow separation on the suction side is identified by the green lines.
Furthermore, for the solid phase, particles represented by brown spheres convect downstream through the passage.
During this process, many particles interact with transitional and turbulent flow structures and are modulated by flow unsteadiness.
For example, on the suction side, particles in case d1 are able to interact with vortices shed from the separation region, as shown in Fig.~\ref{xy-overview}$(a)$.
As the Stokes number increases, particle inertia gradually dominates, resulting in less directional variation of particle velocity \cite{Barker2012Coal}.
Therefore, fewer particles enter the suction-side boundary layer, as indicated by the wider particle-free (blank) regions in Figs.~\ref{xy-overview}$(b)$ and \ref{xy-overview}$(c)$.
In comparison, the pressure-side boundary layer is populated with particles across all cases, suggesting more frequent particle-blade interactions.

\subsection{Particle-Blade Collisions and Model-Based Predictions of Deposition and Erosion}

\subsubsection{First-Collision Locations and Frequencies}
\label{Collision_Locations_and_Frequencies}

To characterize particle-blade interactions, we first examine the spatial distribution and frequency of particle collisions with the blade.
All collision, deposition, and erosion statistics reported in this study are restricted to the first collision of each particle with the blade surface.
These first-collision data are determined before application of the idealized rebound condition and are therefore independent of the modeled post-impact trajectory.
Repeated-impact statistics are not considered because most particles are predicted to deposit after their first collision (see the discussion in \S~\ref{Particle_Deposition}). Moreover, the reflection condition might introduce additional uncertainties.

\begin{figure}[!t]
	\centering\small
	\begin{overpic}[width=0.33\textwidth]{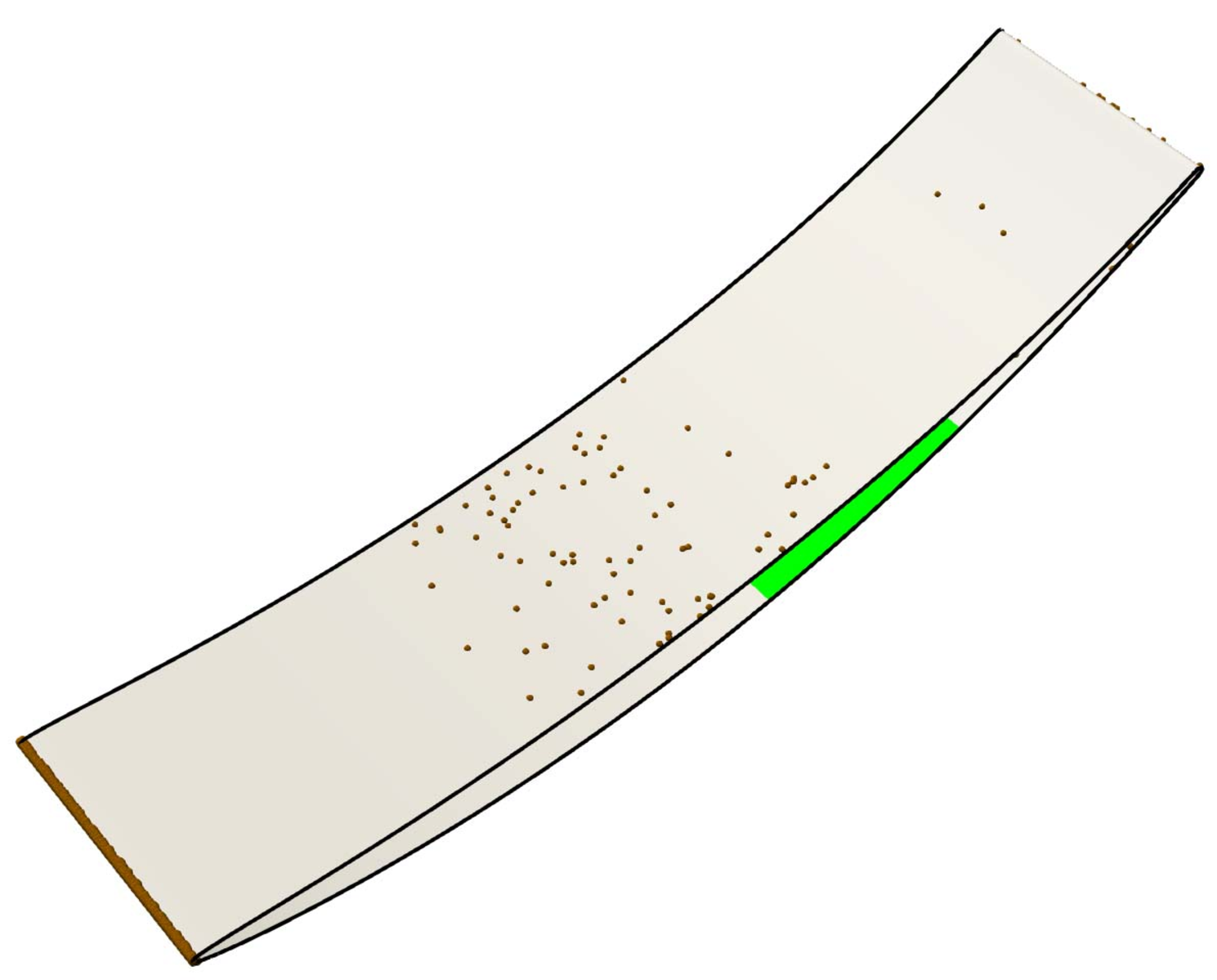}
		{
			\put(0,70){$(a)$}
		}
	\end{overpic}
	\begin{overpic}[width=0.33\textwidth]{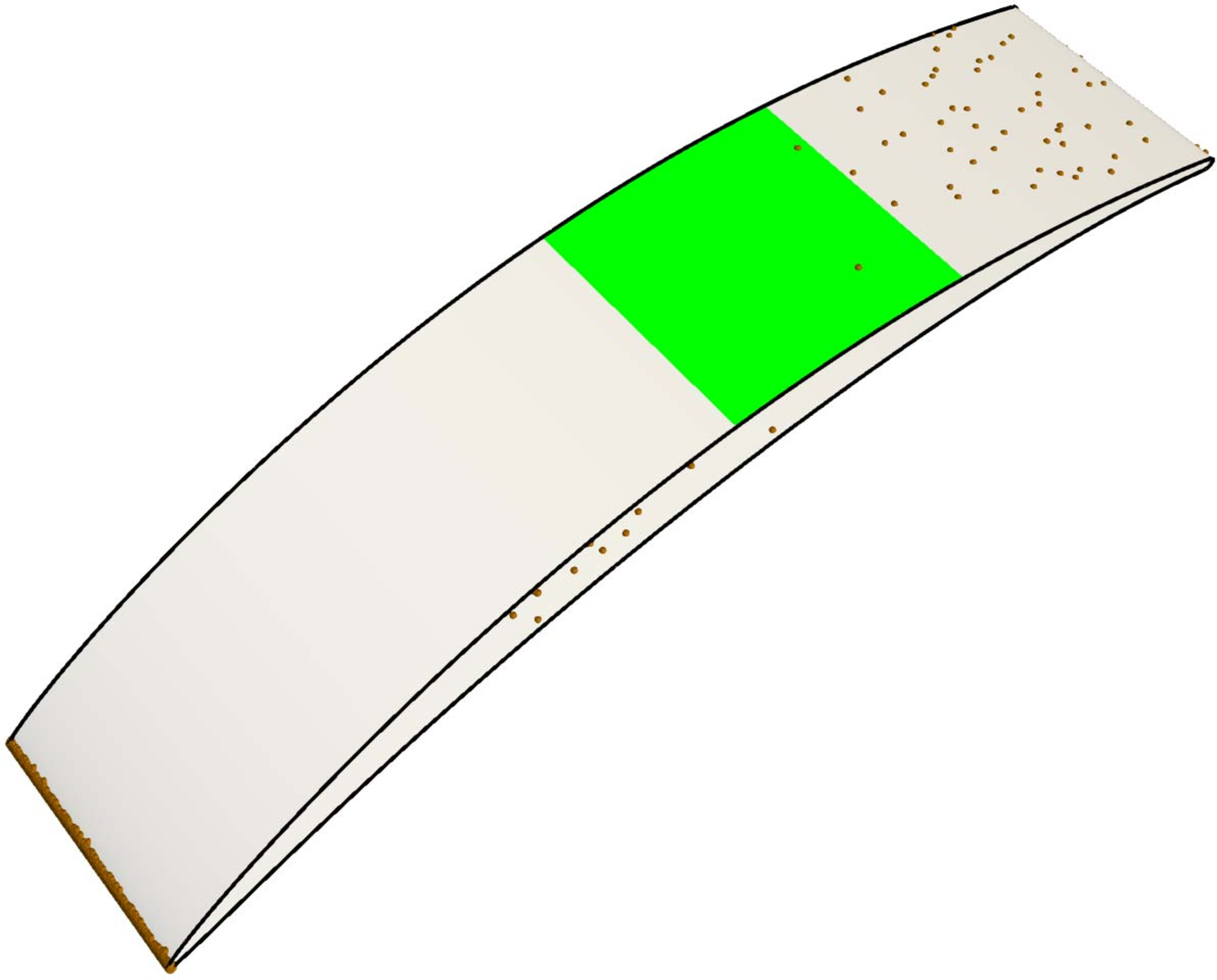}
		{
			\put(0,70){$(b)$}
		}
	\end{overpic}
	\begin{overpic}[width=0.33\textwidth]{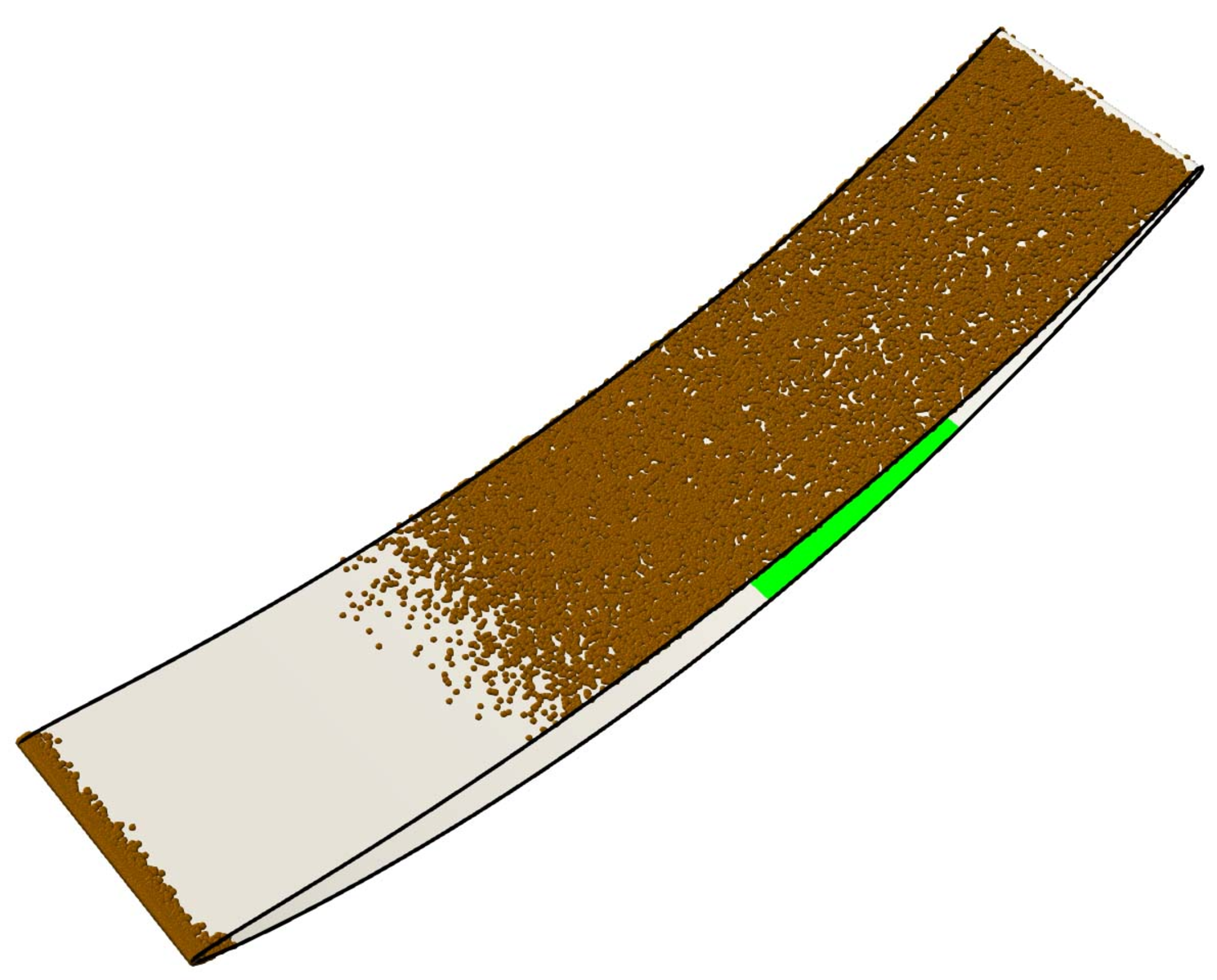}
		{
			\put(0,70){$(c)$}
		}
	\end{overpic}
	\begin{overpic}[width=0.33\textwidth]{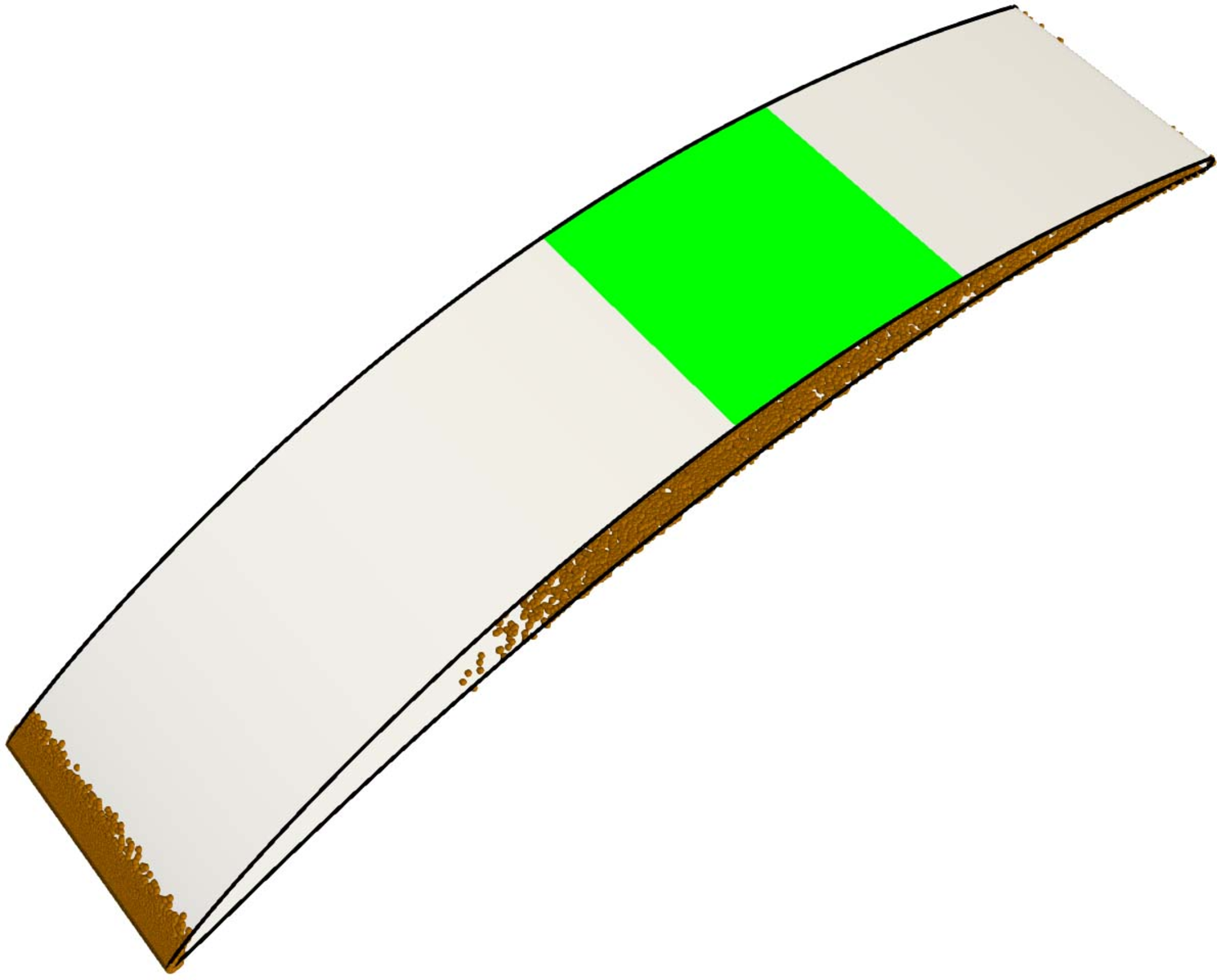}
		{
			\put(0,70){$(d)$}
		}
	\end{overpic}
	\begin{overpic}[width=0.33\textwidth]{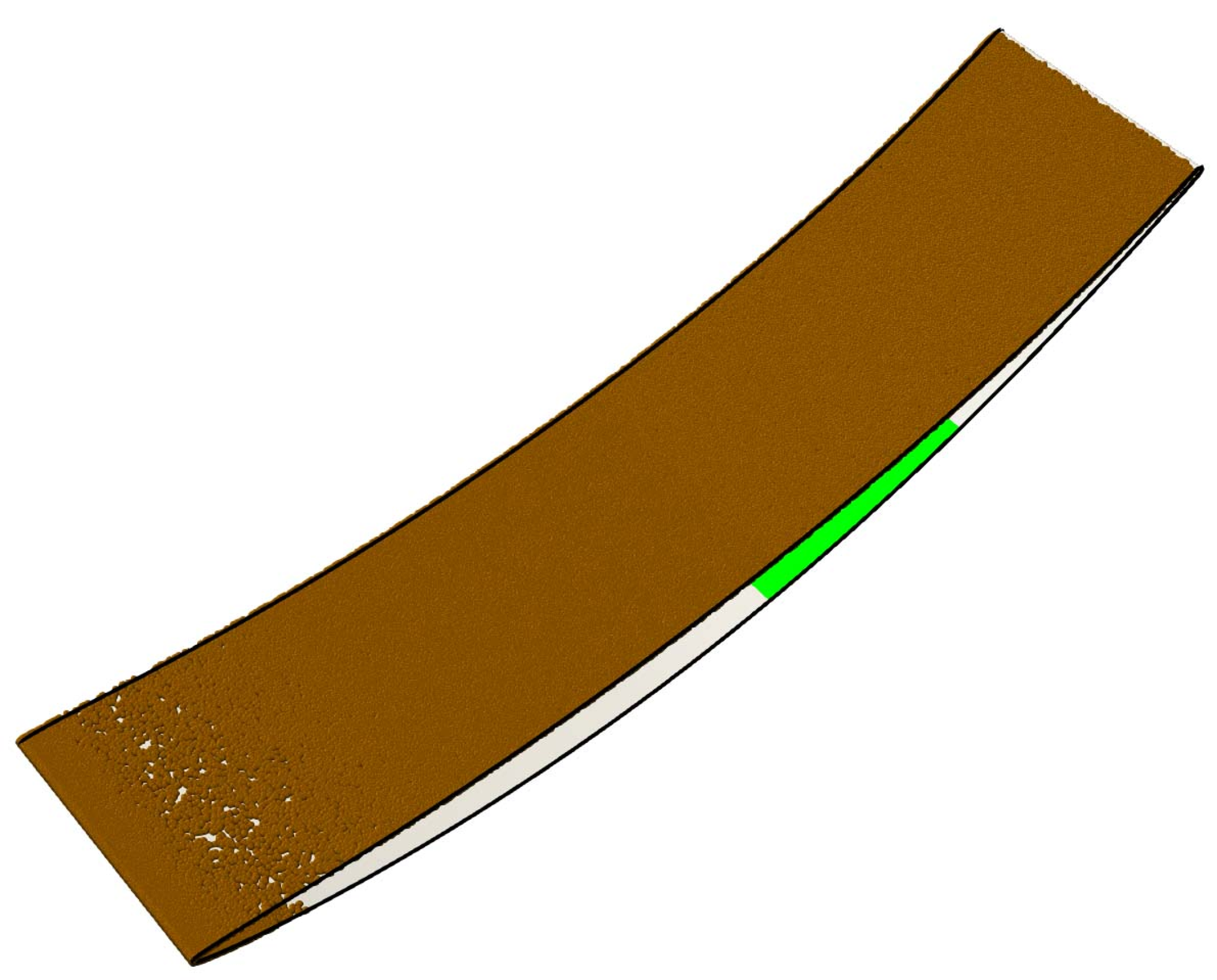}
		{
			\put(0,70){$(e)$}
		}
	\end{overpic}
	\begin{overpic}[width=0.33\textwidth]{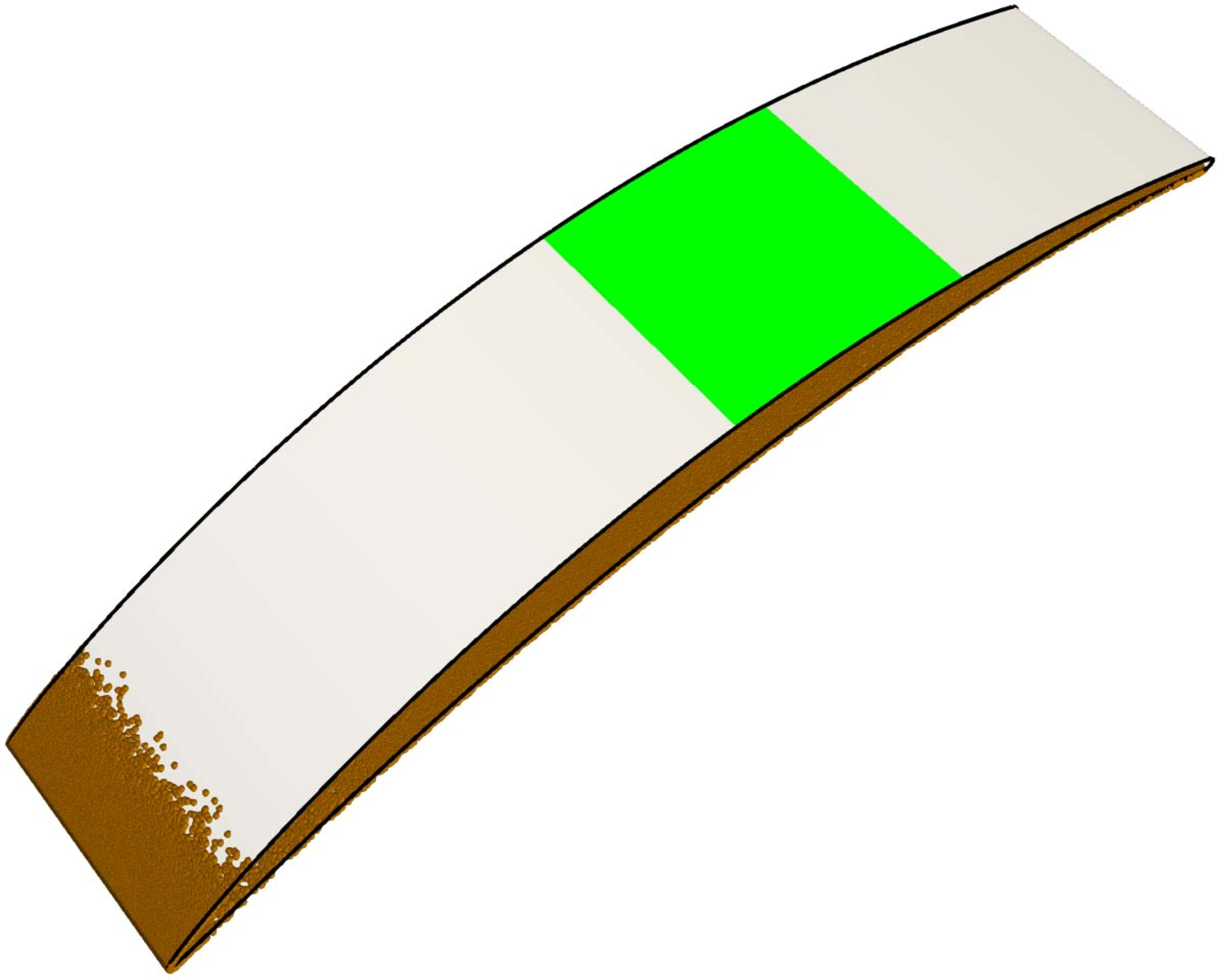}
		{
			\put(0,70){$(f)$}
		}
	\end{overpic}
	\caption{Particles at their first recorded impact with the blade surface for $(a,b)$: case d1; $(c,d)$: case d2; $(e,f)$: case d4.
	The three left panels show the pressure side, whereas the three right panels show the suction side. The mean separation region is marked by the green surface, and each brown marker denotes the location of one particle's first recorded impact.}
	\label{collision-overview} 
\end{figure}

First, we examine the spatial distribution of collisions over the blade surface, and the results are presented in Fig.~\ref{collision-overview}. 
We can see that the vicinity of the blade leading edge experiences frequent particle impacts, and the affected region expands downstream on both sides as the Stokes number increases.
On the pressure side, collisions in case d1 are mainly scattered within the mid-chord region, as shown in Fig.~\ref{collision-overview}$(a)$.
For case d2, shown in Fig.~\ref{collision-overview}$(c)$, a clear line divides the impacted and non-impacted zones, and this boundary is located near the onset region of transition. Further details are provided in \S~\ref{Particle_Responses_to_Flow_Unsteadiness}.
The number of collision events is significantly higher than that in case d1.
Moreover, collisions in case d4 occur over the entire pressure side as presented in Fig.~\ref{collision-overview}$(e)$, which is presumably because the particles with large inertia show ballistic-like trajectories \cite{Barker2012Coal}.
On the suction side, sparse collisions are observed downstream of the mean separation region, as exhibited in Fig.~\ref{collision-overview}$(b)$.
Notably, this phenomenon does not occur in the other two cases in the same region, consistent with the absence of particles revealed in Figs.~\ref{xy-overview}$(b)$ and \ref{xy-overview}$(c)$.

\begin{figure}[!t]
	\centering\small
	\begin{overpic}[width=0.49\textwidth]{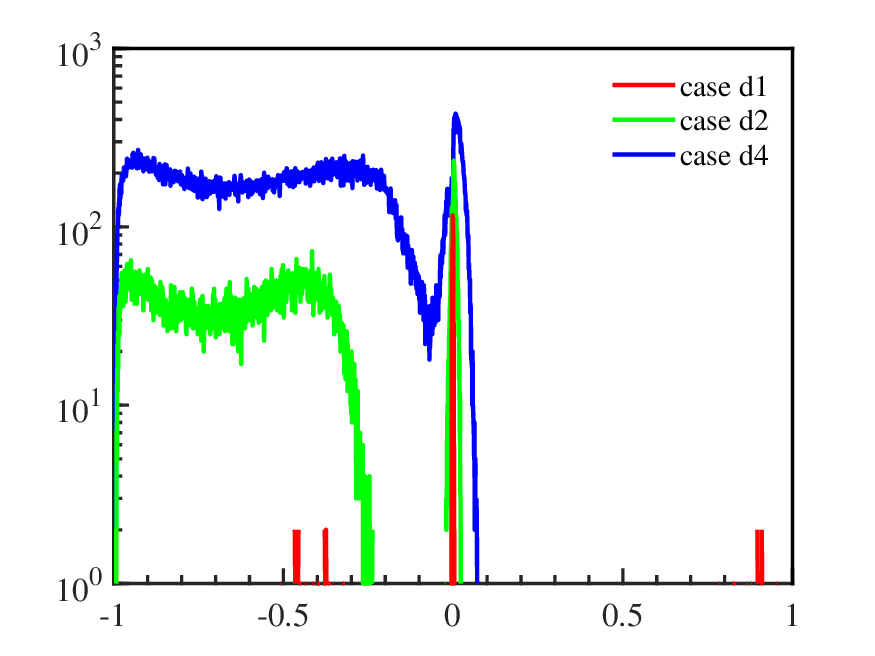}
		{
			\put(3,38){$N$}
			\put(50.5,0){$x_s$}
		}
	\end{overpic}
	\caption{The axial distribution of the number of first-collision events. Here, negative $x_s$ values denote the pressure side, whereas positive $x_s$ values denote the suction side.}
	\label{collision_collect_number}
\end{figure}
\begin{table}[!b]
	\caption{Total number of first-collision events for each case.}
	\begin{center}
		\label{collision_events}
		\begin{tabular}{p{50pt}p{40pt}}
			\hline\hline
			\centering Case &
			\multicolumn{1}{c}{Number of collision events} \\
			\hline
			\centering case d1 &
			\multicolumn{1}{c}{$9645$} \\
			\centering case d2 & 
			\multicolumn{1}{c}{$46298$} \\
			\centering case d4 & 
			\multicolumn{1}{c}{$191035$} \\
			\hline\hline
		\end{tabular}
	\end{center}
\end{table}

To further quantify the particle-blade collisions, we present the axial distribution of the number of collision events in Fig.~\ref{collision_collect_number}.
To distinguish the two blade surfaces in the following surface-distribution plots, a signed axial coordinate $x_s$ is introduced, with $x_s=-x<0$ on the pressure side and $x_s=x>0$ on the suction side.
We can observe that the collision count near the leading edge exhibits a peak, which broadens as the Stokes number increases, corresponding to the enlargement of the impacted region at the leading edge as shown in Fig.~\ref{collision-overview}.
On the pressure side, as the Stokes number increases, the number of collision events increases significantly.
Notably, case d4 exhibits a high collision count over the entire pressure-side surface, while the collision count of case d2 exhibits a sudden increase at $-x_s\approx0.25$.
Moreover, case d1 shows visible particle-blade collisions in limited regions at $0.35\lesssim -x_s\lesssim0.45$, suggesting that the transitional flow structures in this region may enhance the wall-normal transport of low-inertia particles toward the blade.
On the suction side, collisions are mainly detected in case d1 downstream of $x_s\approx0.9$ and the collision count is very low, while no collisions are observed in the other two cases.
Moreover, the total number of first-collision events for each case is summarized in Table~\ref{collision_events} to facilitate the assessment of statistical reliability.

\subsubsection{Particle Responses to Flow Unsteadiness}
\label{Particle_Responses_to_Flow_Unsteadiness}

\begin{figure}[!t]
	\centering\small
	\begin{overpic}[width=0.49\textwidth]{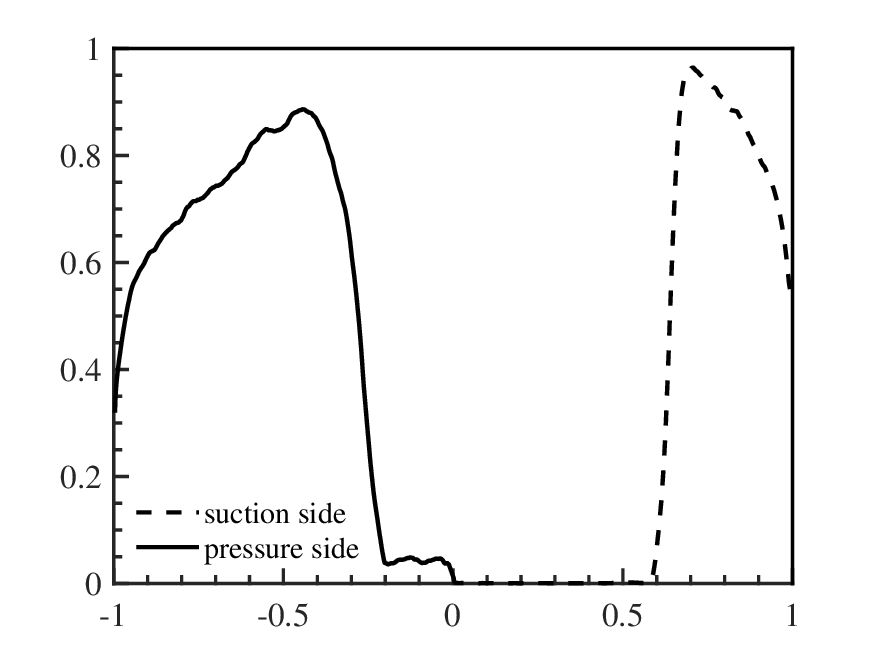}
		{
			\put(3,38){$\gamma_{\max}$}
			\put(50.5,0){$x_s$}
		}
	\end{overpic}
	\caption{Peak intermittency around the boundary layer. Here, negative $x_s$ values denote the pressure side, whereas positive $x_s$ values denote the suction side.}
	\label{intermittency}
\end{figure}

To elucidate the underlying mechanisms for the different collision distributions in the present cases, a detailed investigation of particle response to flow unsteadiness is undertaken.
Based on the different transition mechanisms on the pressure and suction sides \cite{Zaki2010Direct}, the flow unsteadiness is divided into two types: one is the bypass transition to turbulence on the pressure side, and the other is the unsteady vortex shedding induced by flow separation on the suction side.

The boundary-layer transition can be characterized by intermittent flow structures that develop into turbulence.
This behavior can be quantified using the intermittency factor $\gamma$, which represents the probability that the local flow is turbulent.
Following the discrimination algorithm proposed by Nolan and Zaki \cite{Nolan2013Conditional}, the calculation procedure is summarized as follows.
First, a detector function $D$ is defined as the sum of absolute values of wall-normal and spanwise velocity fluctuations, namely $D=|u^\prime_{\mathrm{f},n}|+|u^\prime_{\mathrm{f},z}|$.
Then, a local standard deviation filter is applied to the $D$ field using a three-dimensional stencil \cite{Marxen2019Turbulence}.
Moreover, at each wall-normal location $n$, a threshold for the filtered data is determined using Otsu's method \cite{Otsu1979threshold}.
A location is classified as turbulent when the filtered value exceeds this threshold, corresponding to a logical indicator function $\varGamma=1$; otherwise, it is classified as non-turbulent and $\varGamma=0$. 
Subsequently, the mean intermittency $\gamma$ is obtained by averaging $\varGamma$ over the spanwise direction and time,
\begin{equation}
	\gamma(x_s,n)=\frac{1}{TL_z}\int_{0}^{L_z}\int_{t_0}^{t_0+T}\varGamma\mathrm{d}t\mathrm{d}z.
\end{equation}
Furthermore, the maximum intermittency in the wall-normal direction is evaluated, 
\begin{equation}
	\gamma_{\max}(x_s)=\max_n(\gamma),
\end{equation}
and the results are shown in Fig.~\ref{intermittency}.

\begin{figure}[!t]
	\centering\small
	\begin{overpic}[width=0.49\textwidth]{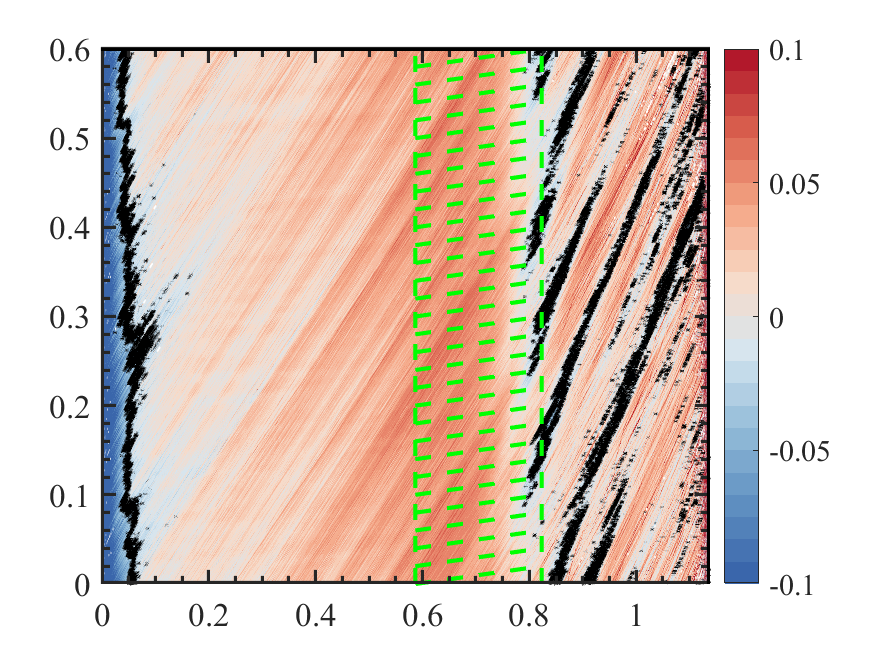}
		{
			\put(0,68){$(a)$}
			\put(3,38){$t$}
			\put(91,38){$\langle u_{\mathrm{p},n}\rangle$}
			\put(48,0.5){$s$}
		}
	\end{overpic}
	\begin{overpic}[width=0.49\textwidth]{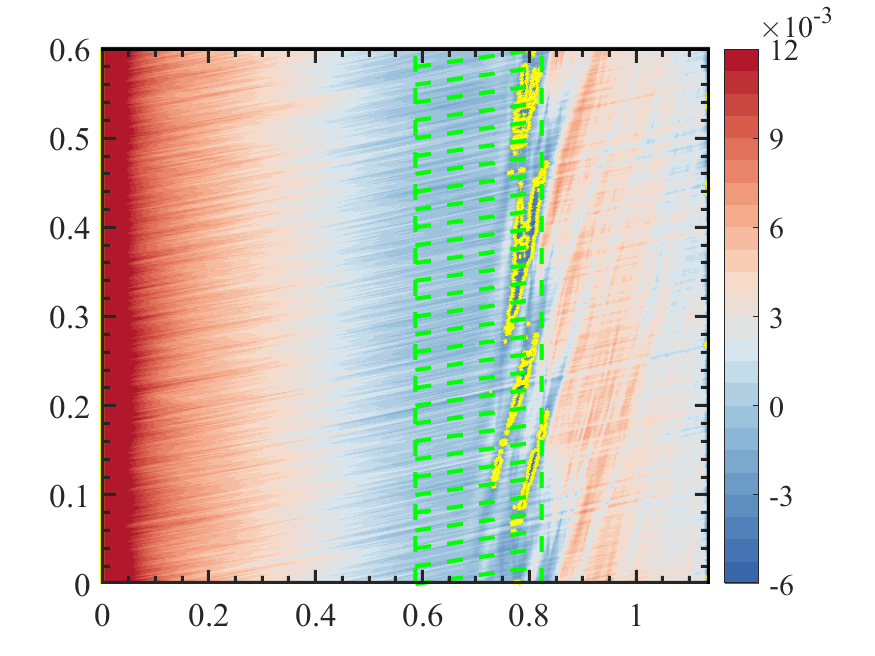}
		{
			\put(0,68){$(b)$}
			\put(3,38){$t$}
			\put(91,38){$\langle C_f\rangle$}
			\put(48,0.5){$s$}
		}
	\end{overpic}
	\caption{Temporal and spatial distribution of the spanwise-averaged $(a)$ particle wall-normal velocity and $(b)$ instantaneous wall-friction coefficient along the blade surface, where $s$ represents the arc-length from the leading edge.
		Particle data are sampled from a layer normal to the blade surface over $n\in[0.02,0.04]$.
		In both panels, the regions filled by green dashed lines indicate the mean separation area.
		In addition, the black lines in Fig.~\ref{response}$(a)$ denote the contour $\langle u_{\mathrm{p},n}\rangle=-0.03$, whereas the yellow lines in Fig.~\ref{response}$(b)$ denote the contour $\langle C_f\rangle=-0.003$.}
	\label{response}
\end{figure}

Based on the intermittency plot, the pressure side is divided into three regions: a laminar region $(0<-x_s\lesssim0.2)$, a transitional region $(0.2\lesssim -x_s\lesssim0.45)$, and a turbulent region $(0.45\lesssim -x_s<1)$.
The pronounced increase in the case-d2 collision count shown in Fig.~\ref{collision_collect_number} begins at $-x_s\approx0.25$, within the transitional interval.
Moreover, particles with a small Stokes number in case d1 show collisions predominantly within the range $(0.35\lesssim -x_s\lesssim0.45)$, where $\gamma_{\max}$ is high.
This spatial correspondence indicates an association between pressure-side collision activity and transition-related intermittency, though it does not establish a causal relationship.
Nevertheless, for case d4, this spatial correspondence is less evident because particles with larger inertia tend to follow ballistic trajectories \cite{Barker2012Coal}, and the particle-blade collisions occur over the entire pressure-side surface.
On the suction side, the intermittency shows an increase at $0.6\lesssim x_s\lesssim0.7$ in Fig.~\ref{intermittency}.
As previously discussed, collisions on this side occur only in case d1, and appreciable collision activity begins at $x_s\approx0.9$, lagging behind the transition onset $x_s\approx0.6$.
This discrepancy suggests that the collisions are not directly caused by the transition process itself.
Instead, they are presumably caused by the response of particles to intense separation-induced vortex-shedding dynamics.

Moreover, to examine the particle response to separation-induced vortex shedding events, we compare the temporal and spatial evolution of the spanwise-averaged wall-normal particle velocity $\langle u_{\mathrm{p},n}\rangle$ and wall-friction coefficient $\langle C_f\rangle$, as shown in Figs.~\ref{response}$(a)$ and \ref{response}$(b)$, respectively.
Downstream of the mean separation region, particles undergo periodic up-and-down movements as they travel downstream, indicated by alternating stripes of positive and negative values in the $\langle u_{\mathrm{p},n}\rangle$ contour plot.
Notably, the contour $\langle C_f\rangle=-0.003$ in Fig.~\ref{response}$(b)$ exhibits a qualitatively similar space-time pattern, with a comparable frequency and propagation slope.
This space-time correspondence suggests modulation of the particle motion by separation-induced vortex shedding.
The negative wall-normal-velocity contours show wall-directed particle transport in the same region, and some of these particles subsequently collide with the blade surface.

\subsubsection{Deposition Criterion and Predicted Spatial Distribution}
\label{Particle_Deposition}

As discussed above, particle-blade interactions can result in either particle adhesion to the blade surface or rebound after impact.
Following the modeling framework of Casari \textit{et al.} \cite{Casari2018EBFOG}, adhesion is treated as deposition, whereas erosion is attributed exclusively to rebounding particles.

To determine whether an impacting particle deposits on the blade surface, the deposition model proposed by Bons \textit{et al.} \cite{Bons2017Simple}, which has been commonly used in turbomachinery deposition studies, is adopted.
In this model, the normal coefficient of restitution, $\mathrm{CoR}_n$, is introduced and defined as the ratio of the rebound normal velocity $u_{\mathrm{p},n2}$ to the impact normal velocity $u_{\mathrm{p},n1}$.
The particle is considered to deposit on the surface when $\mathrm{CoR}_n = 0$, whereas it rebounds from the surface when $\mathrm{CoR}_n > 0$.
It is assumed that the particle initially undergoes elastic compression and then undergoes plastic deformation only if the yield condition is satisfied, and $u_{\mathrm{p},n2}$ is determined by
\begin{equation}
\frac{1}{2}m_{\mathrm{p}}u_{\mathrm{p},n2}^2=
\left\{
    \begin{aligned}
		&\frac{1}{2}m_{\mathrm{p}}u_{\mathrm{p},n1}^2-W_{\mathrm{cont}} & \frac{1}{2}m_{\mathrm{p}}u_{\mathrm{p},n1}^2\leq E_{\mathrm{crit}}\\
        &E_{\mathrm{crit}}-W_{\mathrm{cont}} & \frac{1}{2}m_{\mathrm{p}}u_{\mathrm{p},n1}^2>E_{\mathrm{crit}}
	\end{aligned}
    \right.,
\end{equation}
where $E_{\mathrm{crit}}=\sigma^2_yV_{\mathrm{p}}/(2E_c)$ is the maximum stored elastic energy.
$\sigma_y$ is the yield stress, and $E_c$ is the composite modulus defined as 
\begin{equation}
    \frac{1}{E_c}=\frac{1-\nu_\mathrm{p}^2}{E_\mathrm{p}}+\frac{1-\nu_\mathrm{b}^2}{E_\mathrm{b}},
\end{equation}
where $E_\mathrm{p}$ and $E_\mathrm{b}$ are the Young's moduli of the particle and the blade, respectively, and $\nu_\mathrm{p}$ and $\nu_\mathrm{b}$ denote the corresponding Poisson's ratios.
Here, $W_{\mathrm{cont}}=\gamma_s A_{\mathrm{cont}}$ represents the work required to overcome adhesion forces \cite{Johnson1971Surface}, where $\gamma_s$ denotes the surface free energy and is typically taken as $0.8~\mathrm{J/m^2}$ \cite{Bons2017Simple}, and $A_{\mathrm{cont}}$ is the contact surface area given by
\begin{equation}
	\dfrac{A_{\mathrm{cont}}}{A_{\mathrm{crit}}}=a+b\left(\frac{w_{\mathrm{max}}}{w_{\mathrm{crit}}}\right)^c.
\end{equation}
Here, $a=0.1$, $b=1/7$, and $c=0.5$ are constants.
$A_{\mathrm{crit}}=V_{\mathrm{p}}/(l_{\mathrm{p}}-w_{\mathrm{crit}})$ is the surface area at maximum elastic deformation $w_{\mathrm{crit}}=\sigma_yl_{\mathrm{p}}/E_c$, where $l_{\mathrm{p}}=2d_{\mathrm{p}}/3$ is the equivalent length.
Here, $w_{\mathrm{max}}$ denotes the maximum deformation
\begin{equation}
	w_{\mathrm{max}}=l_{\mathrm{p}}-\exp\left[\ln(l_{\mathrm{p}}-w_{\mathrm{crit}})-\frac{m_{\mathrm{p}}u_{\mathrm{p},n1}^2/2-E_{\mathrm{crit}}}{\sigma_yV_{\mathrm{p}}}\right].
\end{equation}

\begin{figure}[!t]
	\centering\small
	\begin{overpic}[width=0.49\textwidth]{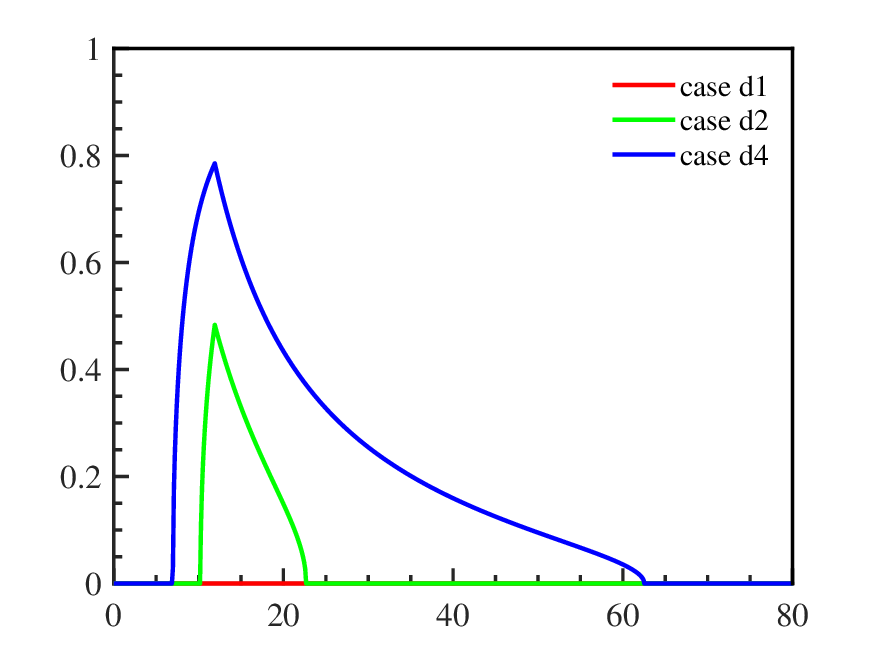}
		{
			\put(1,38){$\mathrm{CoR}_n$}
			\put(45,0){$u^*_{\mathrm{p},n1}$ (m/s)}
		}
	\end{overpic}
	\caption{Normal coefficient of restitution $\mathrm{CoR}_n$ versus dimensional impact normal velocity $u^*_{\mathrm{p},n1}$ for each case.}
	\label{CoRn}
\end{figure}

Based on the model of Bons \textit{et al.} \cite{Bons2017Simple}, the predicted normal coefficient of restitution $\mathrm{CoR}_n$ for each particle size is presented in Fig.~\ref{CoRn}.
Because the mechanical and adhesive properties required by the deposition model are not available for the specific CMAS composition considered here, a quartz-particle/aluminum-blade material pair is adopted as a surrogate to estimate deposition propensity.
Quartz is a major constituent associated with CMAS-type particulate matter \cite{Bansal2015Properties}, and the quartz properties required by the model are available in the literature \cite{Grant1975Erosion,Bons2017Simple}.
The adopted parameters are $\sigma_y=120~\mathrm{MPa}$, $E_\mathrm{p}=72~\mathrm{GPa}$, $E_\mathrm{b}=69~\mathrm{GPa}$, $\nu_\mathrm{p}=0.17$, $\nu_\mathrm{b}=0.32$ \cite{Bons2017Simple}.
Under the adopted quartz-on-aluminum surrogate model, $1~\mu\mathrm{m}$ particles are predicted to deposit over the full range of normal impact velocities considered, because the resulting $\mathrm{CoR}_n$ in Fig.~\ref{CoRn} remains zero.
In contrast, for the two larger particle sizes, the model predicts deposition only over specific ranges of normal impact velocity.

Among first-collision events, the Bons criterion \citep{Bons2017Simple} predicts deposition fractions of $100\%$, $85.88\%$, and $80.44\%$ for cases d1, d2, and d4, respectively.
All subsequent deposition and erosion statistics are based exclusively on these first-impact data.
Repeated impacts and post-impact trajectories are not quantified because their prediction would require a physically dissipative rebound model.

\begin{figure}[!t]
	\centering\small
	\begin{overpic}[width=0.49\textwidth]{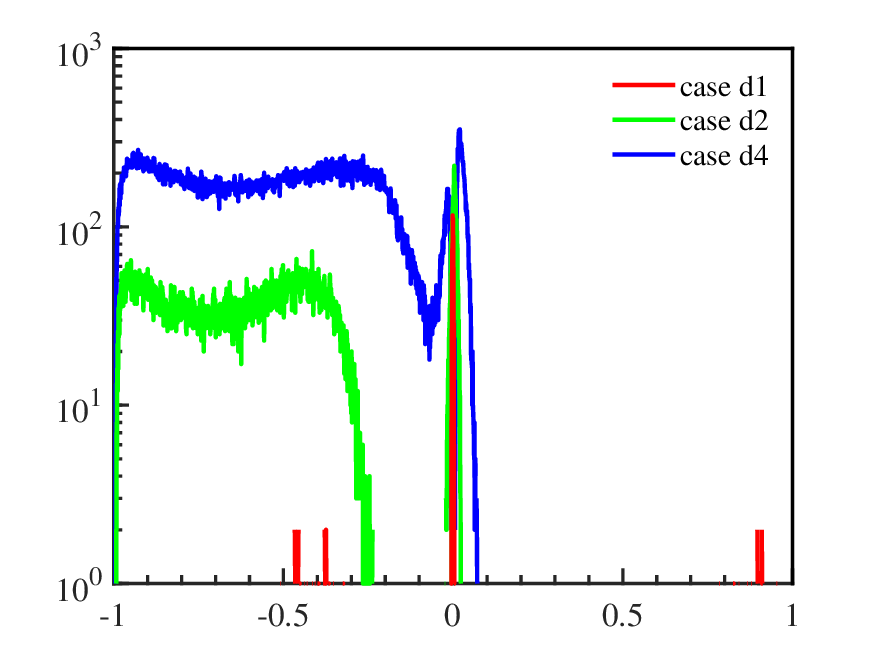}
		{
			\put(3,38){$N$}
			\put(50.5,0){$x_s$}
		}
	\end{overpic}
	\caption{The axial distribution of first-collision events predicted to result in deposition according to the Bons criterion \cite{Bons2017Simple}. Here, negative $x_s$ values denote the pressure side, whereas positive $x_s$ values denote the suction side.}
	\label{deposition_collect_number}
\end{figure}

The axial distribution of particles whose first collision satisfies the Bons deposition criterion \cite{Bons2017Simple} is further presented in Fig.~\ref{deposition_collect_number}.
The distribution closely resembles the corresponding first-collision distribution in Fig.~\ref{collision_collect_number} because most first impacts are predicted to result in deposition.
The predicted deposition pattern is qualitatively consistent with previous experiments, including the deposition near the leading edge and on the pressure side \cite{Tabakoff1987Laser,Dunn2012Operation,Webb2012Coal}, as well as the near absence of deposition on the suction side except close to the leading edge \cite{Smith2010Deposition,Webb2012Coal}.
This comparison constitutes a cross-study consistency check rather than direct validation because the particle properties, surface materials, and operating conditions differ.

Moreover, the correspondence between the first-collision distribution and boundary-layer intermittency, together with the similarity between the collision and predicted deposition distributions, is consistent with a spatial association between transition-related intermittency and increased wall-directed particle transport discussed in \S~\ref{Particle_Responses_to_Flow_Unsteadiness}.
Similar correspondence has also been reported in previous experiments.
For example, Hignett \cite{Hignett1962use} used solid paraffin wax particles, most of which had a diameter of approximately $1.6~\mu\mathrm{m}$, to visualize boundary-layer transition.
The resulting dust deposits were found to delineate the transition region and reveal turbulence wedges downstream of surface protrusions.
This dust deposition technique was subsequently applied to fan blades, where transition lines on the blade surfaces were identified \cite{Hignett1963Surface}.

\subsubsection{Predicted Erosion Distribution}

To assess the blade erosion associated with the first-collision events predicted to result in rebound, we employ the Grant--Tabakoff erosion model \cite{Grant1975Erosion}, which was specifically developed for turbomachinery applications and has been widely used in previous studies \cite{Ghenaiet2012Study,Casari2018EBFOG,Ghenaiet2024Modeling}.

The Grant--Tabakoff erosion model defines the erosion rate $\varepsilon$ as the mass of wall material removed ($\mathrm{mg}$) per unit mass of impacting particles ($\mathrm{g}$), which is given by 
\begin{equation}
	\varepsilon=K_1\left[1+C_k K_{12}\sin\left(\dfrac{\pi/2}{\beta_0}\beta_1\right)\right]^2
    u^{*2}_\mathrm{p1}\cos^2\beta_1[1-R_{t}^{2}] + K_3(u^*_\mathrm{p1}\sin\beta_1)^4.
\end{equation}
Here, $u^*_\mathrm{p1}$ is the particle impact velocity and $\beta_1$ is the impact angle measured from the local wall tangent.
For quartz particles impinging on an aluminum blade surface \cite{Grant1975Erosion}, the model coefficients used in this study are as follows: $K_1=3.67\times10^{-6}$, $K_{12}=0.585$, $K_3=6\times10^{-12}$ and $R_t=1-0.0016u^*_\mathrm{p1}\sin\beta_1$.
Here, $C_k$ is a function of $\beta_1$, which is given by
\begin{equation}
	C_k=
    \left\{
    \begin{aligned}
	    &1 && \beta_1 \leq 2\beta_0\\
        &0 && \beta_1 > 2\beta_0
	\end{aligned}
    \right.,
\end{equation}
where $\beta_0=\pi/9$ is the impingement angle corresponding to maximum erosion.
Moreover, the overall damage to the blade is assessed by the erosion-rate density \cite{Ghenaiet2024Modeling} $E$ ($\mathrm{mg\cdot s^{-1}\cdot mm^{-2}}$), which is defined as 
\begin{equation}
	E=\frac{1}{A^*\Delta t^*}\sum_{i=1}^{N_\mathrm{imp}}m^*_\mathrm{p}\varepsilon_i.
\end{equation}
Here, $N_\mathrm{imp}$ is the number of first-impact events predicted to result in rebound on a given surface element during the time interval of $\Delta t^*$ ($\mathrm{s}$), and $A^*$ is the local surface area ($\mathrm{mm^2}$).
\begin{figure}[!t]
	\centering\small
	\begin{overpic}[width=0.49\textwidth]{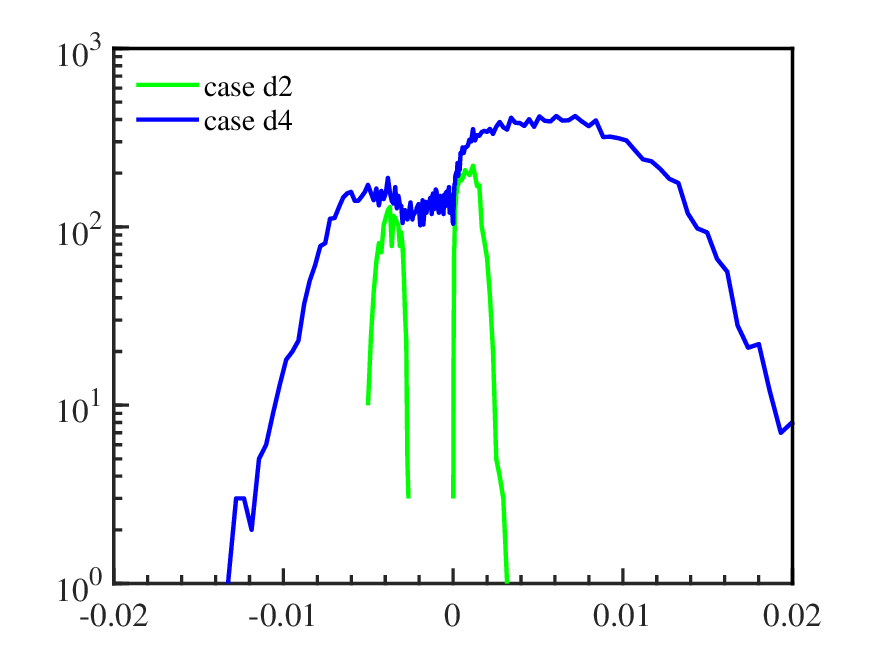}
		{
			\put(3,38){$N$}
			\put(50.5,0){$x_s$}
		}
	\end{overpic}
	\caption{The axial distribution of first-collision events classified as rebound by the Bons criterion \cite{Bons2017Simple}. Here, negative $x_s$ values denote the pressure side, whereas positive $x_s$ values denote the suction side.}
	\label{erosion_collect_number}
\end{figure}
\begin{figure}[!t]
	\centering\small
	\begin{overpic}[width=0.49\textwidth]{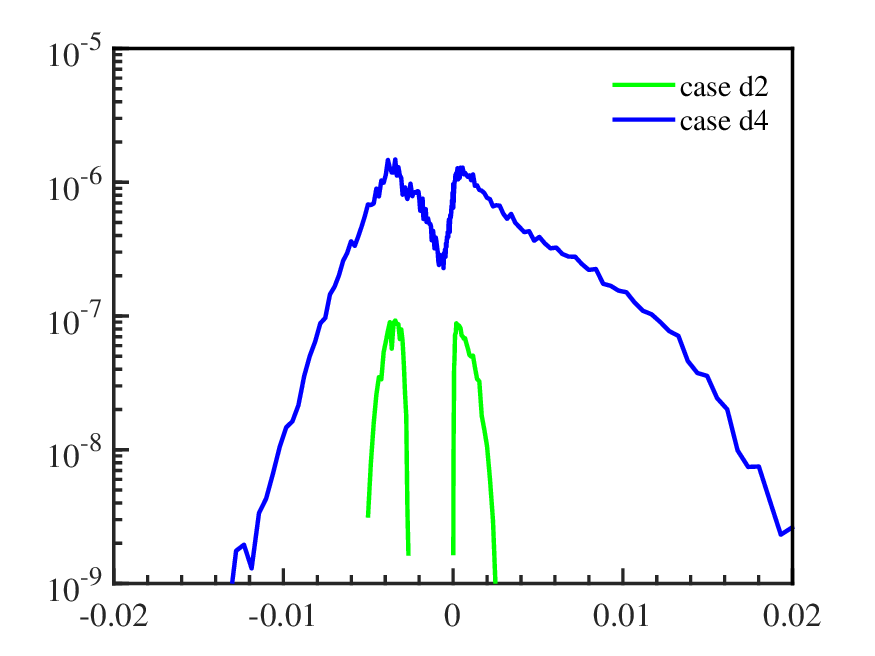}
		{
			\put(1,38){$E$}
			\put(50.5,1){$x_s$}
		}
	\end{overpic}
	\caption{The axial distribution of the model-predicted erosion-rate density $E$ ($\mathrm{mg\,s^{-1}\,mm^{-2}}$), evaluated using first-collision events classified as rebound by the Bons criterion \cite{Bons2017Simple}. Here, negative $x_s$ values denote the pressure side, whereas positive $x_s$ values denote the suction side.}
	\label{erosion} 
\end{figure}

Based on the discussion in \S~\ref{Particle_Deposition}, the adopted modeling framework predicts no erosion for case d1 because all recorded first impacts satisfy the Bons deposition criterion \cite{Bons2017Simple} and are excluded from the erosion calculation.
For cases d2 and d4, both rebound-classified first-impact events and nonzero model-estimated erosion are concentrated near the leading edge, as presented in Figs.~\ref{erosion_collect_number} and \ref{erosion}, respectively.
Because the same particle-number injection rate produces different particle-mass loadings, and because the erosion model involves particle-size extrapolation and surrogate material properties, neither the absolute magnitude of $E$ nor its difference between particle-size cases is interpreted quantitatively.
Nevertheless, the predicted spatial distributions consistently identify the leading edge as the primary erosion-prone region, in qualitative agreement with previous experimental observations \cite{Ghenaiet2004Experimental}.

\begin{figure*}[!t]
	\centering\small
	\begin{overpic}[width=0.32\textwidth]{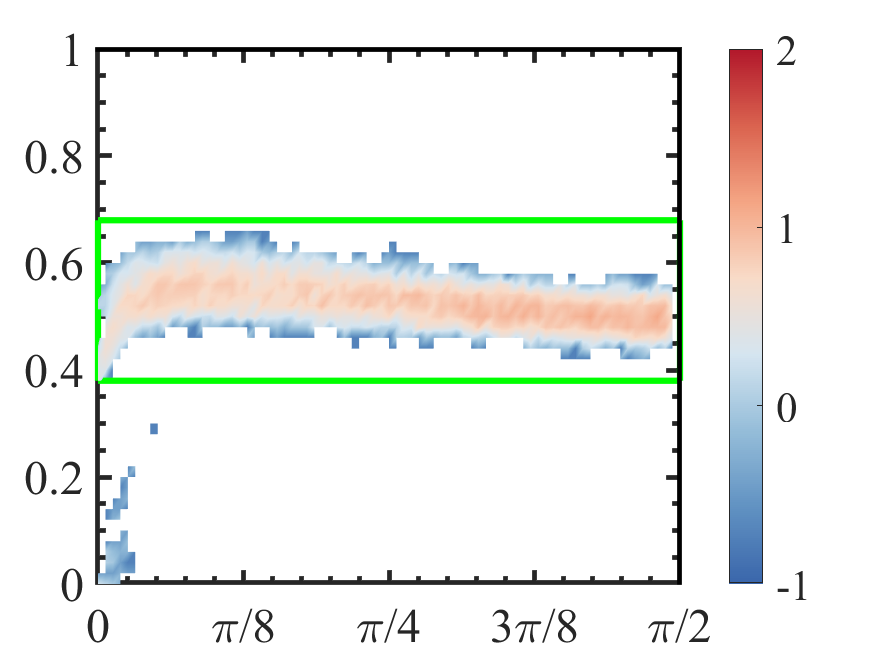}
		{
			\put(-2,68){$(a)$}
			\put(-1,38){$u_\mathrm{p1}$}
			\put(91,49){\rotatebox{-90}{$\log_{10}(C)$}}
			\put(42,-5){$\beta_1$}
		}
	\end{overpic}
	\begin{overpic}[width=0.32\textwidth]{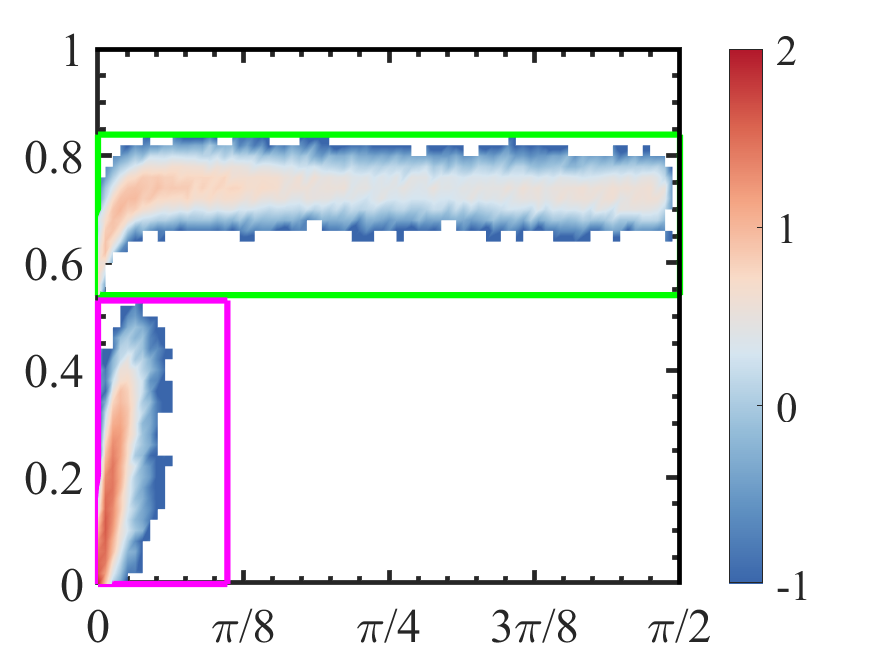}
		{
			\put(-2,68){$(b)$}
			\put(-1,38){$u_\mathrm{p1}$}
			\put(91,49){\rotatebox{-90}{$\log_{10}(C)$}}
			\put(42,-5){$\beta_1$}
		}
	\end{overpic}
	\begin{overpic}[width=0.32\textwidth]{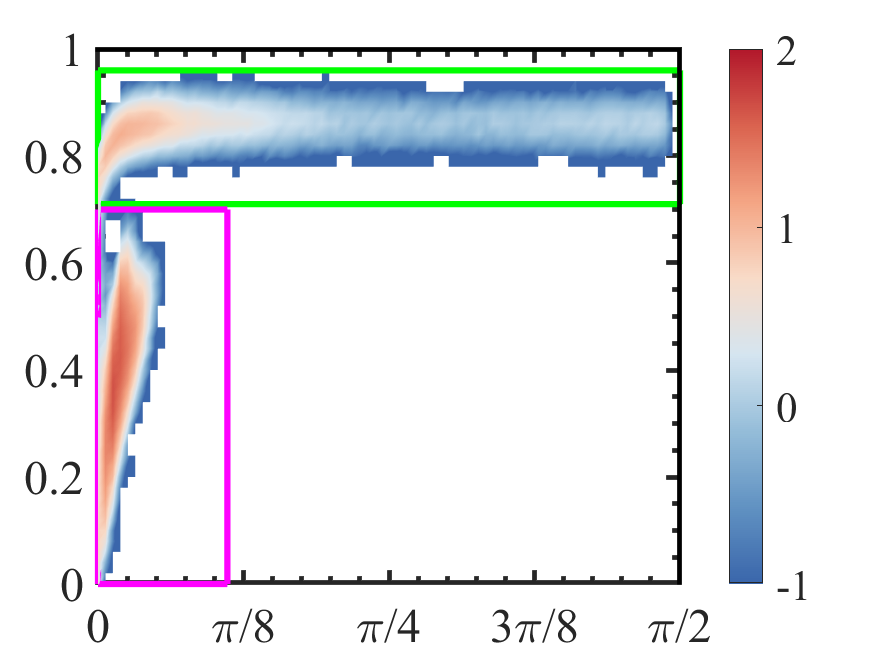}
		{
			\put(-2,68){$(c)$}
			\put(-1,38){$u_\mathrm{p1}$}
			\put(91,49){\rotatebox{-90}{$\log_{10}(C)$}}
			\put(42,-5){$\beta_1$}
		}
	\end{overpic}
	\caption{The joint PDF of particle impact velocity $u_\mathrm{p1}$ and impact angle $\beta_1$ at the first collision for $(a)$: case d1; $(b)$: case d2; $(c)$: case d4.
    Here, $C=p(u_{\mathrm{p1}},\beta_1)$ denotes the normalized joint probability density, satisfying $\int\!\!\int C\,\mathrm{d}u_{\mathrm{p1}}\,\mathrm{d}\beta_1=1$, and the colour scale shows $\log_{10}(C)$.
	In addition, the green boxes mark collision events occurring near the leading edge; magenta boxes mark collisions on the pressure side.}
	\label{PDF}
\end{figure*}

To understand why appreciable erosion is predicted only near the leading edge, we examine the joint probability distribution function (PDF) of the particle impact velocity and angle at the first collision, because both quantities determine the predicted rebound behavior \cite{Bons2017Simple} and enter the Grant--Tabakoff erosion model \cite{Grant1975Erosion}.
These results are presented in Fig.~\ref{PDF}.
For case d1, the PDF predominantly exhibits a horizontal stripe pattern characterized by a nearly constant impact velocity.
This pattern clearly results from collision events occurring near the leading edge, which is marked by the green box.
Moreover, for cases d2 and d4, a distinct oblique stripe pattern marked by the magenta box emerges from the origin, corresponding to a number of additional collision events occurring on the pressure side.
In comparison, collisions near the leading edge involve relatively high impact velocities and span a wide range of impact angles, while the pressure-side collisions exhibit relatively low impact velocities and small impact angles.
These characteristics imply that leading-edge collisions are associated with relatively higher impact normal velocities, while pressure-side collisions correspond to substantially lower impact normal velocities.
Therefore, the present model predicts appreciable erosion primarily near the leading edge, as discussed earlier.

\section{Conclusions}\label{summary_conclusion}

We performed point-particle direct numerical simulations (PP-DNS) of sand-laden flow through a linear compressor cascade subjected to controlled free-stream turbulence, and used the Bons \cite{Bons2017Simple} and Grant--Tabakoff models \cite{Grant1975Erosion} to predict deposition and erosion, respectively.
The DNS reproduces key boundary-layer features reported in the literature, including pressure-side bypass transition and suction-side separation with reattachment, thereby providing a credible baseline for assessing particle-blade interactions.

The resulting collision hotspots are predicted at the leading edge and on the pressure side.
On the pressure side, the onset and distribution of particle-blade collisions are associated with the combined effects of particle inertia and transitional dynamics.
For small Stokes numbers, collision events are scarce upstream and emerge after boundary-layer intermittency increases during bypass transition, suggesting an association between transitional perturbations and near-wall particle transport. 
For intermediate Stokes numbers, the onset of frequent collision events shifts upstream and aligns with the location of rapid growth in intermittency.
For the largest Stokes number examined, particle trajectories become increasingly ballistic, and collision events occur over most of the pressure side. 
On the suction side, rare collisions are observed only for the smallest particles, mostly downstream of the mean separation bubble. 
The qualitatively similar space-time patterns of the spanwise-averaged wall friction and particle wall-normal velocity suggest that separation-induced vortex shedding modulates particle transport. 

Because most first impacts satisfy the Bons deposition criterion \cite{Bons2017Simple}, the predicted deposition distribution largely follows the collision distribution, with hotspots near the leading edge and on the pressure side.
Furthermore, joint distributions of impact velocity and angle differentiate leading-edge and pressure-side impact regimes: leading-edge impacts combine higher speeds with a wide range of angles, whereas pressure-side impacts cluster at lower speeds and shallower angles. 
Together with the adopted deposition and erosion models, these collision characteristics explain why the model-estimated erosion is concentrated near the leading edge.

Overall, the results provide resolved evidence of spatial and temporal associations between unsteady boundary layer dynamics and first-collision patterns, together with their implications for model-predicted deposition.
These insights may enable improved hotspot identification and durability prediction in particle-laden operating environments. 
Moreover, by resolving all turbulent scales, the present study provides high-fidelity reference data that may support future assessment of engineering-scale particle-transport models.

In future work, several issues could be examined in greater detail.
The present simulations include only the drag force.
An a posteriori force evaluation indicates that the Saffman lift is secondary on an ensemble-average basis under the present conditions, although larger local effects for individual near-wall or impacting trajectories cannot be excluded.
The reported distributions are also conditional on uniform inlet injection in the present study. A nonuniform upstream particle concentration could reweight the incoming trajectories and alter the predicted hotspots \cite{Ghenaiet2012Study}.
In addition, deposition is evaluated by applying the Bons criterion \cite{Bons2017Simple} to first-collision events in post-processing.
The present study therefore does not quantify post-impact trajectories or repeated impacts.
Future simulations could incorporate adhesion and dissipative rebound directly into the trajectory solver using a hard-sphere model with velocity-dependent restitution coefficients or a soft-sphere model accounting for particle deformation and finite contact duration.

\begin{acknowledgment}

The authors wish to thank Prof. Richard D. Sandberg for the license of the HiPSTAR solver.
The authors acknowledge Beijing PARATERA Tech Co., Ltd. for providing high-performance computing resources.
This work was supported by the National Natural Science Foundation of China (Grant Nos. 12588201, 12432010, and 12572247).

\end{acknowledgment}

%
% ...

\bibliographystyle{asmems4}

\bibliography{asme2e}





\end{document}